Attilio Sacripanti

# *A Seoi survey for Coaches and Teachers*

# *A Seoi survey for Coaches and Teachers*

By Attilio Sacripanti

Abstract

After a short review of basic biomechanics of Seoi Family and their characteristic, two fast survey of Seoi in judo books and in worldwide researches are performed.

The lot of information collected around the world flow into the competitive application of this family (in Japanese view) or in the most energy convenient technical variation (in Biomechanical view) with the study of the last new solution or variation applied.

An analysis of complementary tools utilized for each component of the Family to increase their effectiveness or refine their application to Ippon follow. Till to the new ways Chaotic applications based both on totally rotational application or reverse application of Seoi Lever System, with the conclusion at the end by a brief overview of Physical and Biomechanical framework connected to Seoi Family, with a more extended study of interaction ( Seoi application to throws).

Conclusion is presented with a comparative evaluation of some remarkable properties all applicable competitive Seoi techniques useful both for coaches and teachers..

Attilio Sacripanti

# *A Seoi survey for Coaches and Teachers*

## *1. Introduction*

It is already well known that biomechanical classification and physiological evaluation are strictly connected.

In physiology starting from the first pioneering works in Japan [1] 57 years ago, judo throwing techniques were ordered in term of energy expenditure. Fig1

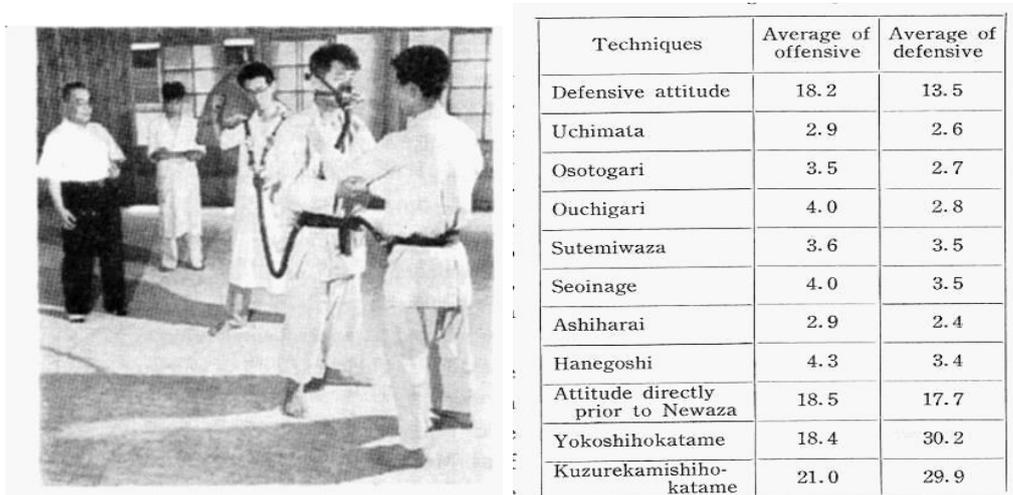

| Techniques | Average of offensive | Average of defensive |
|---|---|---|
| Defensive attitude | 18.2 | 13.5 |
| Uchimata | 2.9 | 2.6 |
| Osotogari | 3.5 | 2.7 |
| Ouchigari | 4.0 | 2.8 |
| Sutemiwaza | 3.6 | 3.5 |
| Seoinage | 4.0 | 3.5 |
| Ashiharai | 2.9 | 2.4 |
| Hanegoshi | 4.3 | 3.4 |
| Attitude directly prior to Newaza | 18.5 | 17.7 |
| Yokoshihokatame | 18.4 | 30.2 |
| Kuzurekamishiho-katame | 21.0 | 29.9 |

*Fig 1-2  First in the world physiological study on Judo in Japan[1]*

The same results were produced by other worldwide researches one of these were performed in recent time in Italy [2]            .

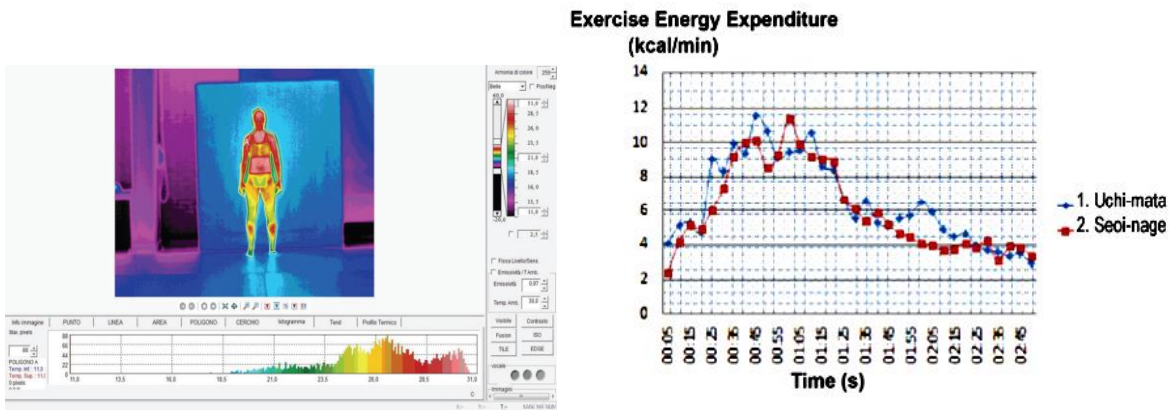

*Fig. 3, 4 Recent physiological study in Italy[2]*

The common underline of these researches implies that all Judo throws can be grouped in two classes.
More Expensive and Less Expensive and these two energy-physiological groups overlap the two groups of Biomechanical classification, which are based on the two tools applied to throws.
It is well know that these two groups, that were single out 28 years ago [3], are the Lever and Couple biomechanical groups.
Till to 2014, among the Lever group the most applied and successful Family of techniques in Japanese view was the Seoi Family, among the couple group Uchi Mata, all this could change in view of refereeing rules that can change during the years.
In this paper we will analyze the Seoi throw and his variation or Seoi Family at first from their pictorial view from historic books, then to consider the different point of view by the results of the researchers in the world about this technique that is the most studied in absolute term.
Follow a short analysis of the complementary tools utilized in high level competition and the evolution of this technique during time.
Starting from the common Physical framework a comparative biomechanical evaluation of the different competitive useful application will be performed.
This paper was shown as presentation at EJU Judo Expert Group during the Second European Symposium: "Science of Judo research" in Antalya, Turkey 05/16/2015.

## 2. *General Biomechanics of Seoi techniques*

The Biomechanics put the Seoi Family in the Lever Group of Biomechanical Classification; this means that the basic mechanics of the throw is the application of a stopping point or fulcrum on the Uke's body applying at the same time a force at a certain distance from the fulcrum (the arm of the lever) that is able to turn the Uke's body around the fulcrum throwing him possibly on the back.
This in short words is the basic mechanics of the Seoi. Fig 5

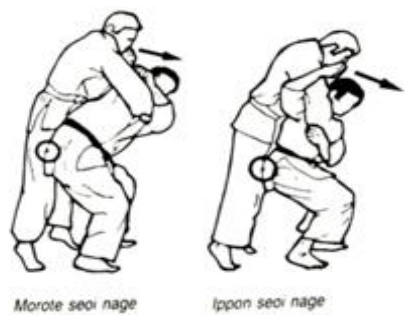

*Fig.5 Basic mechanics of standing Seoi : Lever ( Fulcrum + Force)*

Obviously the energy expenditure is inversely connected to the arm length, or in simpler words longer is the arm of the lever, lesser is the power applied and the energy wasted. Then if the arm of the lever grows less energy is consumed Fig 6,

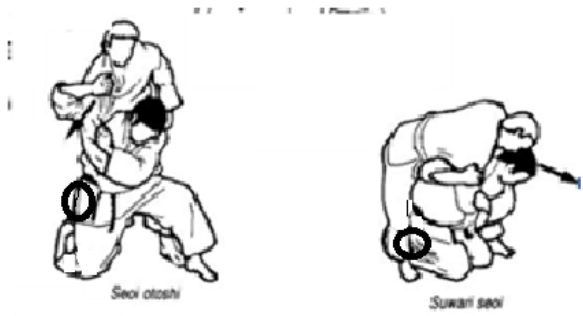

*Fig 6 Seoi Otoshi, and Suwari Seoi two long arm Seoi variation*

In term of biomechanics these two throws are the same lever, with the arm lengthened, but from the Japanese point of view there are three different throws.[4]
Namely Standing Seoi, Seoi Otoshi, Suwari Seoi* (* this last throw name it is not in Japanese tradition but comes from France and Italian habit for *sitting seoi*, for English speaking Drop Seoi). As it easy to see in the next tables: Tab 1, 2.
Both Seoi Nage and Suwari Seoi Nage are among the most utilized and successful judo throws in the world.

| FRA | | JPN | | URSS | | Autres | |
|---|---|---|---|---|---|---|---|
| Uchi-mata | 25,5 % | Uchi-mata | 15,8 % | Uchi-mata | 11,4 % | Suwari-seoi-nage | 13,8 % |
| O-uchi-gari | 11 % | Suwari-seoi-nage | 13,3 % | Seoi-nage, Kata-guruma | 9,6 % | Uchi-mata | 13,4 % |
| O-soto-gari | 7,7 % | Ko-uchi-gari | 10,7 % | Suwari-seoi-nage | 7 % | O-uchi-gari | 8,7 % |
| Sode-tsuri-komi-goshi | 7,4 % | O-uchi-gari | 9 % | Kuchiki-daoshi | 7,7 % | Ko-uchi-gari | 7,4 % |
| Ko-uchi-gari | 7,4 % | O-soto-gari | 7,3 % | O-soto-gari | 7,4 % | Seoi-nage, Kata-guruma | 7,4 % |
| Kuchiki-daoshi | 7,4 % | Tomoe-nage | 6 % | Sode-tsuri-komi-goshi | 7 % | Hara-goshi | 7 % |
| Suwari-seoi-nage | 5,3 % | Seoi-nage, Kata-guruma | 5,1 % | Tai-otoshi | 7 % | | |

*Tab. 1 % of utilization of judo throws by athletes from France Japan and Russia [5]*

| | FRA | JPN | URSS | Autres |
|---|---|---|---|---|
| 1 | Uchi-mata | Suwari-seoi-nage | Seoi-nage, Kata-guruma | O-soto-gari |
| 2 | Kuchiki-daoshi | Seoi-nage, Kata-guruma | Uchi-mata | Suwari-seoi-nage |
| 3 | Seoi-nage, Kata-guruma | Sode-tsuri-komi-goshi | Ura-nage | Sode-suri-komi-goshi |
| 4 | O-uchi-gari | Ko-soto-gari-gaké | O-uchi-gari | O-uchi-gari |
| 5 | O-soto-gari | Tomoe-nage | Sode-tsuri-komi-goshi | Tai-otoshi |
| 6 | Hiza, Sasae | O-soto-gari | Maki-komi | Tomoe-nage |

*Tab 2 Throws effectiveness for athletes from France, Japan and Russia [5]*

Biomechanical analysis shows other interesting findings about Seoi throws.
Generally speaking iss impossible to perform Seoi techniques without unbalance, carefully all the seoi techniques that are kuzushi ( unbalance)  dependent need also to stop for a moment the adversary motion to be carried out.
These techniques are arranged by very complex movements, that in term of Action Invariants  (In Biomechanics *Action Invariants*, should be recognized as the minimum path, in time, of body's shift to acquire the best Kuzushi- Tsukuri position for every Judo Throws) are composed by GAI+SSAI+ISAI+ Lever  [6].
In term of Metabolic Cost they are among the most expensive techniques, and athletes that manage this kind of judo throws are normally provided with high body coordination, that is very important to apply high final acceleration to the adversary body.
.

## 3. *Seoi Family in books*

The problems with the historic research on Seoi Family are essentially connected both on the shape of the throws and on the names , about names it is interesting to reread what Kazuzo Kudo the last Kano student , wrote in his book , Dynamic Judo:
"*Jūdō names fall into the following categories:*

1. *Name that describe the action:* ***ō-soto-gari, de-ashi-barai, ō-uchi-gari-gaeshi.***
2. *Names that employ the name of the part of body used:* ***hiza-guruma, uchi-mata.***
3. *Names that indicate the direction in which you throw your opponent:* ***yoko-otoshi, sumi-otoshi.***
4. *Names that describe the shape the action takes:* ***tomoe-nage*** *('tomoe' is a comma-shaped symbol).*
5. *Names that describe feeling of the techniques:* ***yama-arashi, tani-otoshi.***

*Most frequently, jūdō techniques names will use the content of one or two of these categories.*"
(From: Kudō Kazuzō: *Dynamic Judo Throwing technique* [7]. Appendix p. 220).
 In the paper of Shigeoka  [8]  on the historic development of Seoi it is shown a Seoi  of 1913
 Fig. 7  Historically speaking in the old *Kodokan judo* [ 9] the Seoi Family was composed by Seoi basic technique ( Morote variation, Ippon Seoi Nage , and Seoi Otoshi, with a kneeling position). In *Kanon of judo* [10] by Mifune the Seoi family is presented without specific name in three variation, today known as Morote Seoi Nage, Ippon Seoi Nage, and Eri Seoi Nage. However in this book it is possible to find also other members as Uchi Makikomi, Fig.8, Ganseki Otoshi Fig.9 and Seoi Otoshi ( kneeling), fig 10

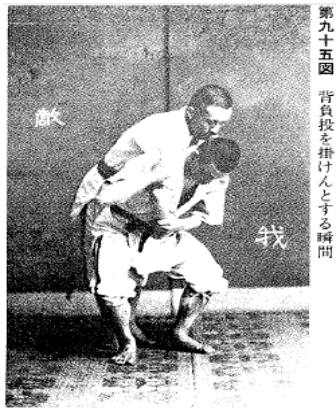

*Fig 7  Ippon Seoi Nage 1913 [8]*

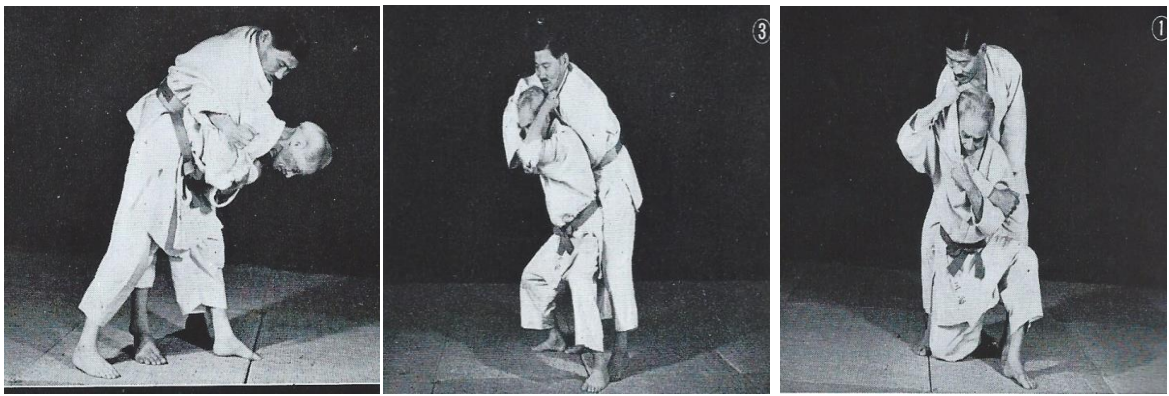

*Fig. 8,9,10 Uchi Makikomi, Ganseki Otoshi, Seoi Otoshi  [10]*

In the book of Mikinosuke Kawaishi , *Ma Methode de Judo* [11]  the Seoi Family is grouped into the Kata Waza, and Ippon Seoi Nage is presented as Kata Seoi, and his left variation as Hidari Kata Seoi, but as new entry we can find  Seoi Age. Fig 11

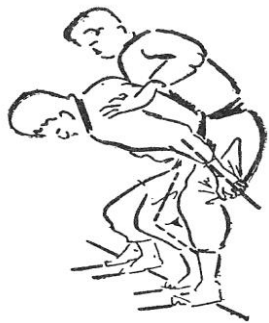

*Fig 11 Seoi Age [11]*

Other contributions on Seoi Family come from the book of the founder of British Judo, Gikin Koizumi in *My Study of Judo* [12] the author present a larger Family with:
Ippon Seoi Nage Fig.12, Eri Seoi Fig.13, Seoi Otoshi kneeling Fig.14, Hiza Seoi Fig.15, Obi Seoi Fig.16, and at last Suso Seoi.Fig. 17.

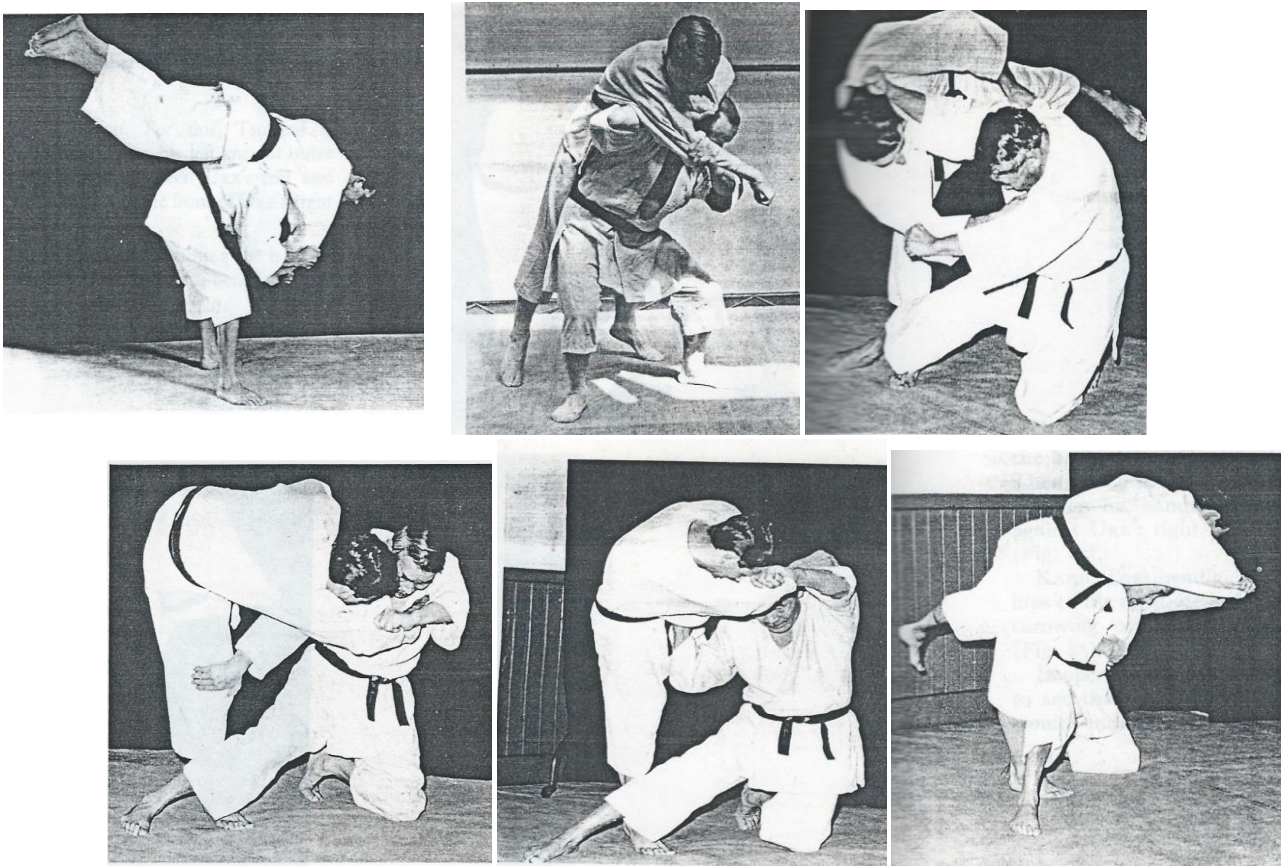

*Fig 12-17 Ippon Seoi Nage, Eri Seoi, Seoi Otoshi, Hiza Seoi, Obi Seoi, Suso Seoi . [12]*

In the Golden book *Dynamic Judo* [13] by Kazuzo Kudo the basic technique Seoi Nage ( Morote Variation) is connected with his continuation Seoi Makikomi , ( called Uchi Makikomi in Mifune) and Ippon Seoi Nage Fig 18-22 ( Hidari Variation ) connected with a different kind of Seoi Otoshi Fig 23-24.

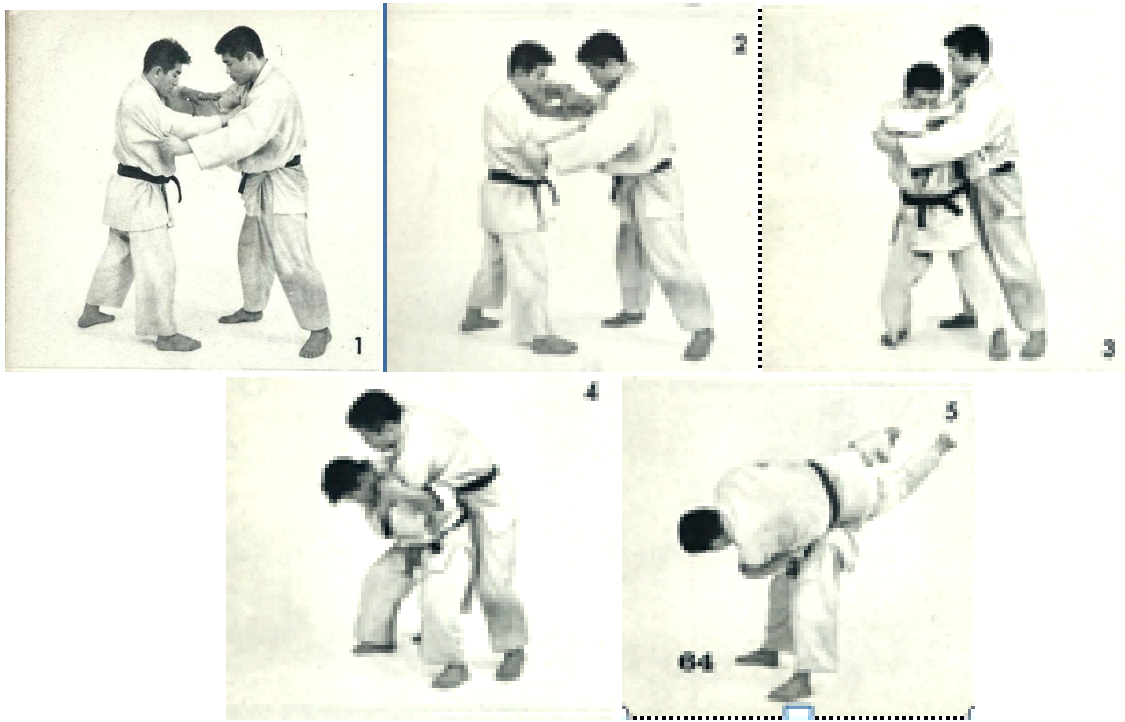

*Fig 18-22 Hidari Ippon Seoi Nage*

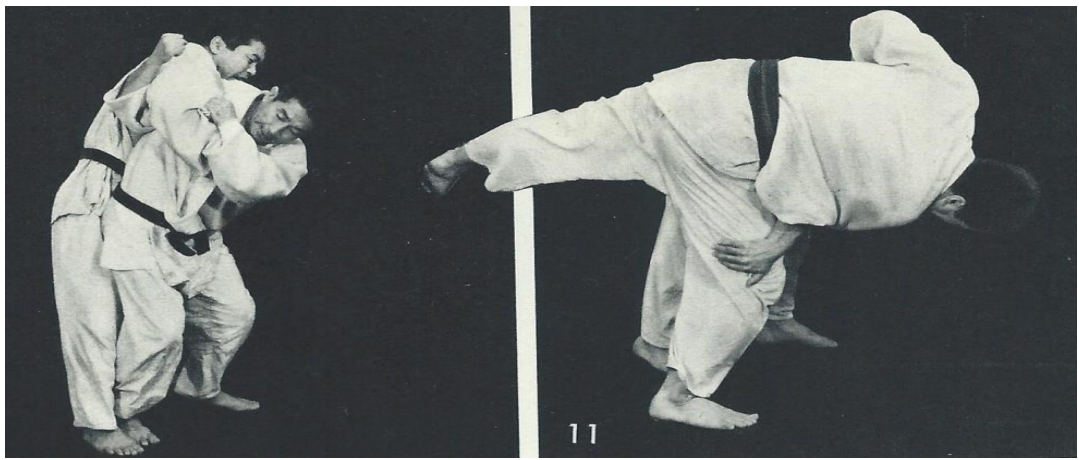

*Fig.23 Seoi Otoshi ( Kudo, Sato) [13,14]*

In *Best Judo* [14] by Inokuma and Sato the basic throw is Ippon Seoi shown in three variations the second one is Seoi Otoshi fig 23 like in the book *Vital Judo* [15] by Sato and Okano, the third one is Morote Seoi, but follow in combination paragraph named Ippon Seoi Nage, both Seoi Otoshi kneeling and Suwari Seoi seated ( exactly equal to the point of view of Biomechanics: same tool with arms of different lengths )
Into the Roy Inman *English Syllabus* [16] the Seoi Family reaches his maximum development presenting a lot of new entries.
Namely Ippon Seoi Nage, Morote Seoi Nage, Kata Sode Seoi Otoshi, Morote Eri Seoi Nage, Ryo Hiza Seoi Otoshi, Seoi Otoshi, Kata Sode Seoi Otoshi, Soto Mata Seoi Otoshi, Uchi Makikomi, Fig.24-31.

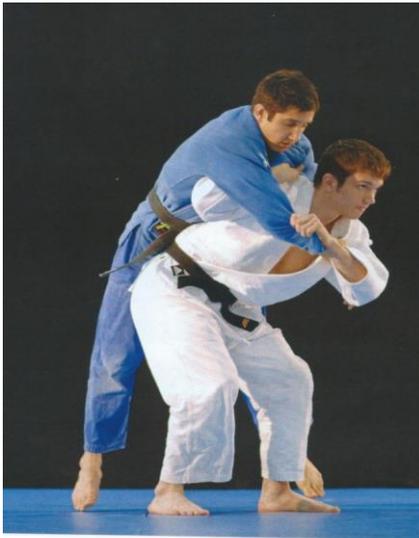
MOROTE-SEOI-NAGE *Two Hand Shoulder Throw*

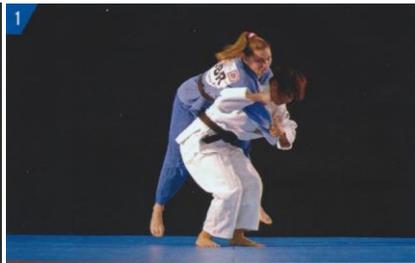
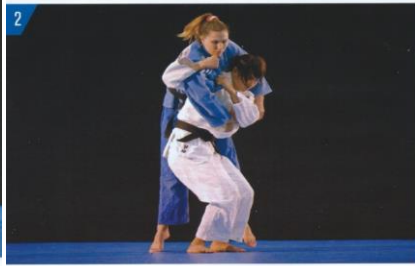
IPPON-SEOI-NAGE *One Arm Shoulder Throw*

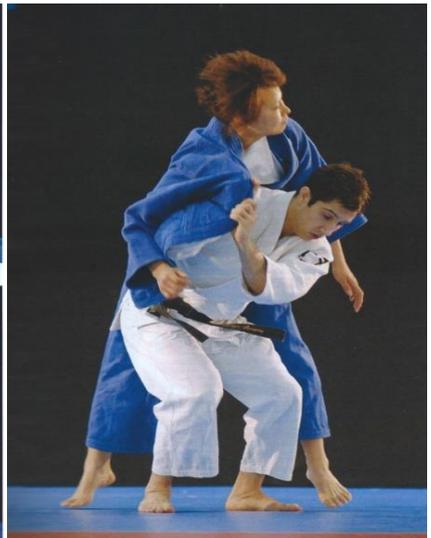
MOROTE-ERI-SEOI-NAGE *Two-handed Lapel Shoulder Throw*

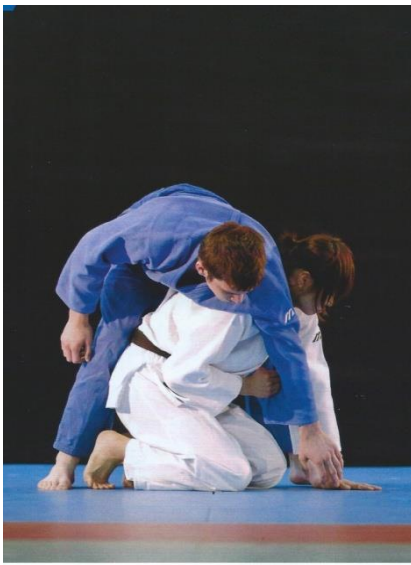
KATA-SODE-SEOI-OTOSHI *Single Sleeve Shoulder Drop*

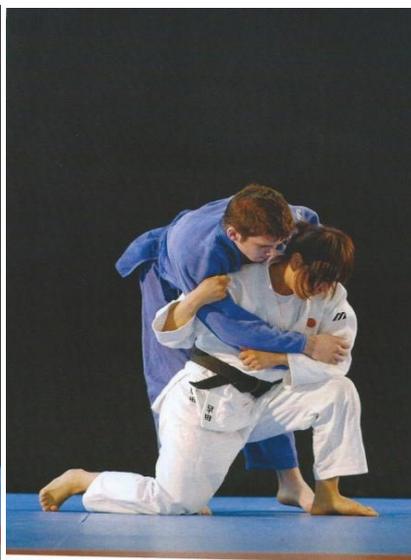
SEOI-OTOSHI *Shoulder Drop*

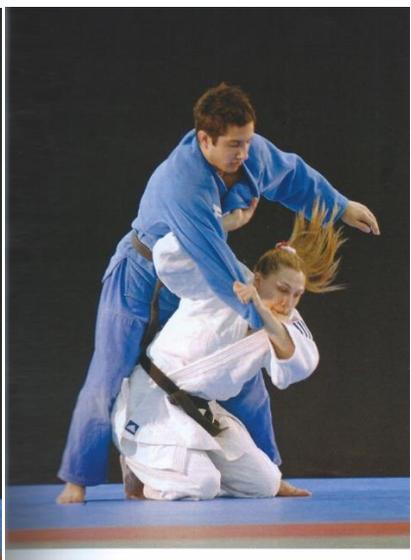
RYO-HIZA-SEOI-OTOSHI *Two Knee Shoulder Drop*

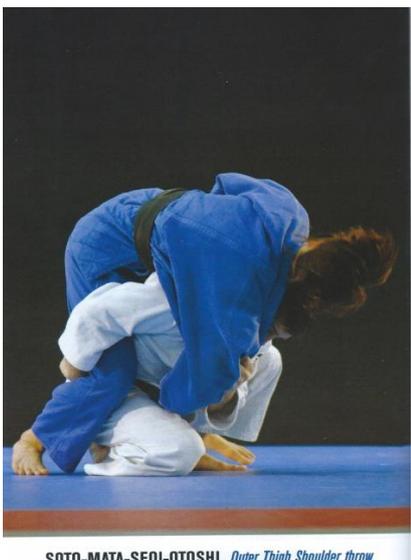
SOTO-MATA-SEOI-OTOSHI *Outer Thigh Shoulder throw*

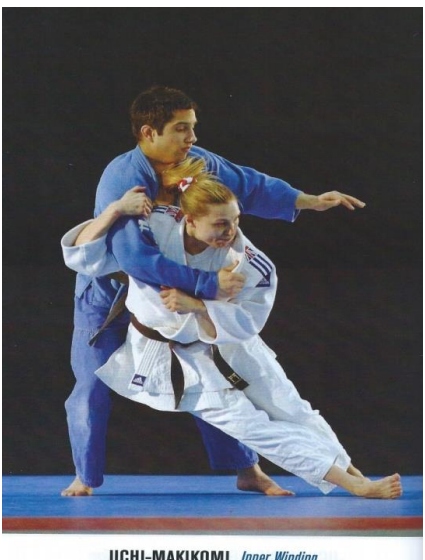
UCHI-MAKIKOMI *Inner Winding*

*Fig. 24-31 Seoi Family in English Syllabus [16]*

In the last *Kodokan Judo throwing techniques* [17] by Toshiro Daigo the Seoi Family changes, many names disappear we find Ippon Seoi Nage Fig 32-34 with five variations the second one is not named, but it is : Seoi Otoshi as in fig 23, also Morote Seoi is named Ippon Seoi Nage with five variation , in which the last one is named Ganseki Otoshi Fig 35-37.

But as new entry as technical autonomous we find Seoi Otoshi kneeling with four variations among them Suwari Seoi, and at last Uchi Makikomi with four variations but put among Yoko Sutemi Fig 38-40.

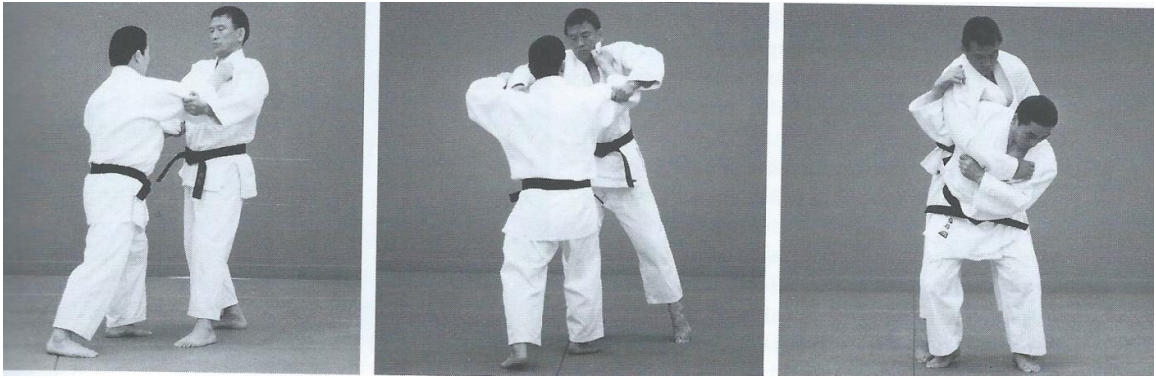

*Fig 32-34 Ippon Seoi [17]*

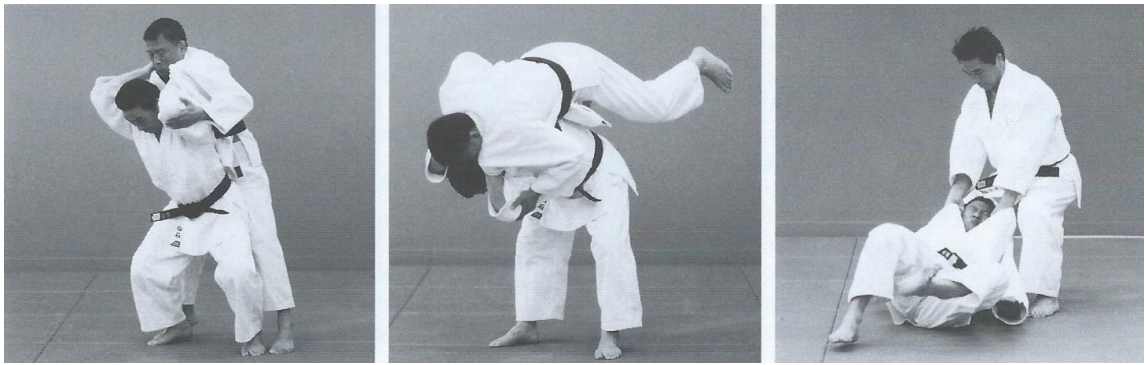

*Fig35-37 Ganseki Otoshi [17]*

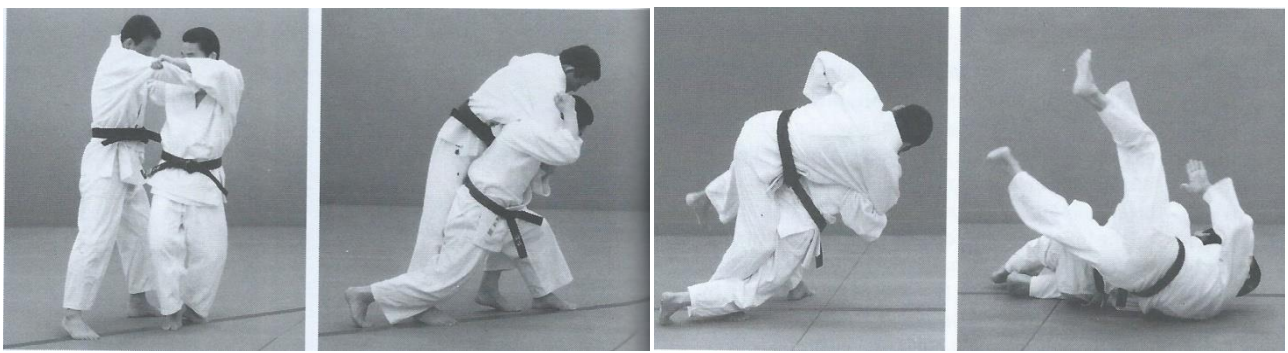

*Fig. 38-41 Uchi Makikomi [17]*

This short overview of Seoi Family in the books shows us the lack of understanding of the mechanics of throws and the problem of throw names among authors.
However today in high level competition only four variation of Seoi Family are applied, with some interesting evolution versus chaotic forms.

## 4. Researches on Seoi : The Worldwide studied technique

The great importance of Seoi in high level competition and his high effectiveness in real contest is the main reason of the lot of studies performed around the world in these last sixty years.
In this paragraph an overview of some worldwide researches will be developed to show how many details and from how many different points of view these techniques were studied in so many universities' laboratories.
Many researches are focalized on the understanding of the mechanics of throws, but if the mechanics could be clear to researchers the books show us, that communication between scientific and sportive areas are deficient.
So many different field of researches show also that Judo Sport, belonging to the class of dual situation sports, is a very difficult and multi-complex task.
The first ever tentative to understand the kinetic of Seoi come obviously from Japan:
Ikai and Matsumoto , *The kinetics of Judo* [18] 1958 (!)  it is possible to see in the next Fig.42-43 the motion of Center of Mass of Athletes and the attack angle between CoM

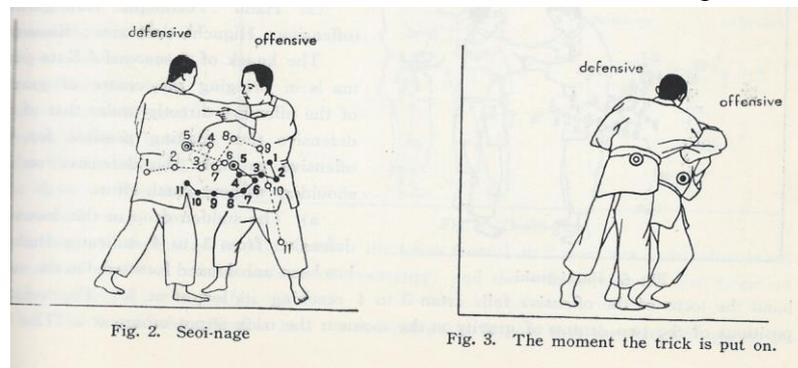

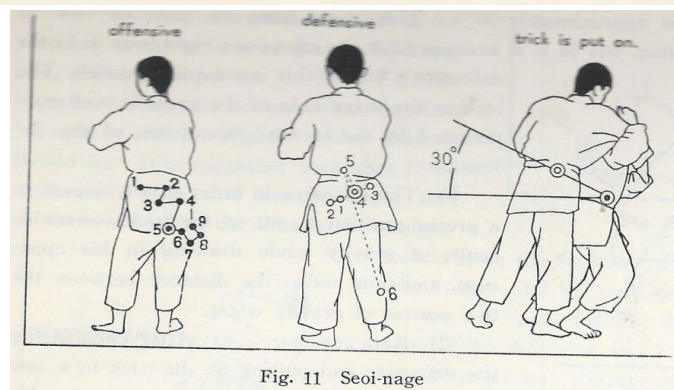

*Fig.42-43 The kinetic of Judo.  (1958)*

The following studies choose among the worldwide most interesting works all focalized on Seoi Nage and his complexities are arranged for Country alphabetic order and not for scientific area.

*Austria*

Hassmann and coworkers : *Judo performance test using a pulling force device simulating a Seoi Nage throws* [19] 2011 Fig.44

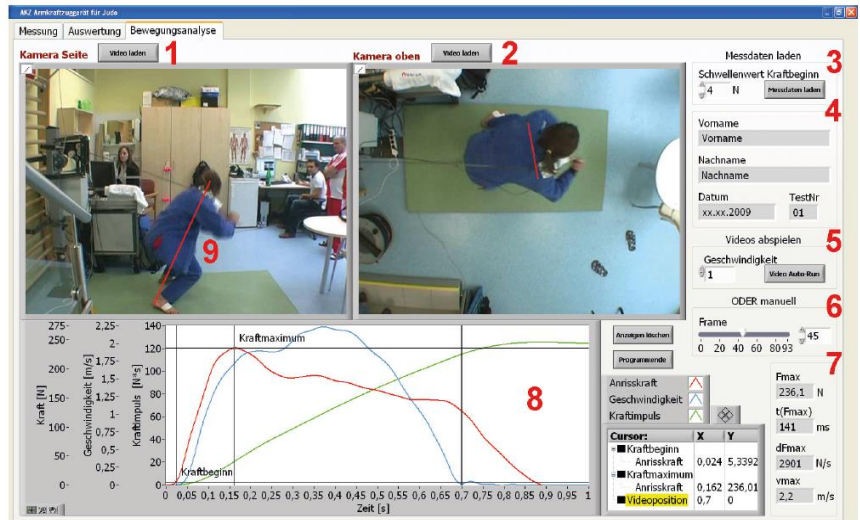

*Fig.44 Layout of the performance test simulating Seoi Nage*

*Brasil*

Lopes Melo and Coworkers : *Cinematica da variacao angular de tronco, quadril e johelo do atacante na tecnica Seoi Nage no Judo* [20] 2004 fig. 45-47

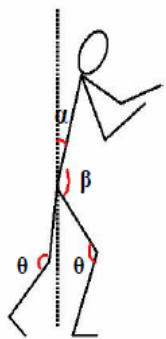 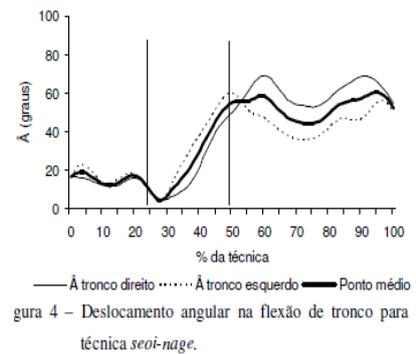

*Fig.45-47 Angular variation of Tori body's segment in Seoi Nage throw*

Franchini and coworkers : *Energy expenditure in different judo throwing techniques* [21] 2008 Fig 48-49

Table 1: Oxygen uptake, heart rate and blood lactate responses for three judo throwing techniques.

|  | Seoi-nage | Harai-goshi | O-uchi-gari |
|---|---|---|---|
| $VO_2$ (mL.kg$^{-1}$.min$^{-1}$) | 33.71±5.68 ** | 32.28±5.10 | 29.97±6.10 |
| Heart rate (bpm) | 146±14 | 140±11 | 139±16 |
| $[La^-]_{rest}$ (mmol.L$^{-1}$) | 0.94±0.26 | 1.02±0.51 | 0.90±0.33 |
| $[La^-]_{peak}$ (mmol.L$^{-1}$) | 1.80±0.56 | 2.02±1.33 | 1.73±0.88 |

$VO_2$ = arithmetic mean of oxygen uptake during 5-min activity for each technique; $[La^-]_{rest}$ = blood lactate concentration before each technique; $[La^-]_{peak}$ = highest blood lactate value measured 1, 3 and 5 min after the completion of 5-min of activity for each technique (one throw each 15s); ** significantly different from *o-uchi-gari* ($P < 0.01$)

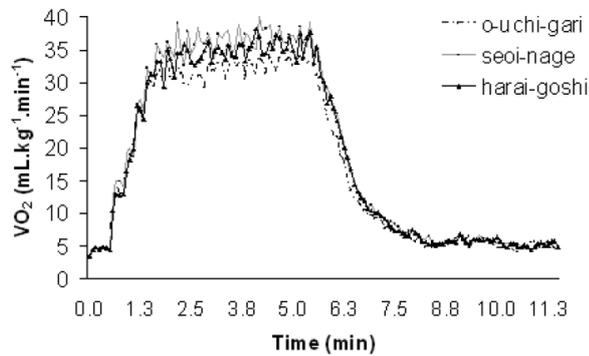

*Fig.48-49 Results confirming the high energy expenditure of Seoi throw [21]*

**Egypt**
Ibrahim Fawzi Mustafa, *Force impulse of body parts as function for prediction of total impulse and performance point of Ippon Seoi Nage skill in judo* [22] 2010 Fig.50-51

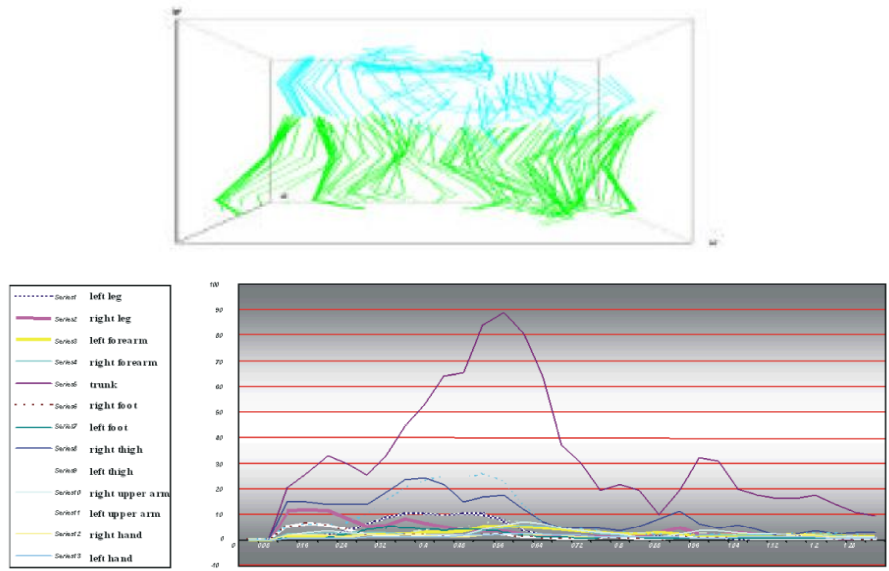

*Fig.50-51 Graph of partial forces and total impulse of Seoi performance.[22]*

**France**

Blais, Trilles : *Analyse méchanique comparative d'une meme projection de Judo: Seoi Nage, realisée par cinq experts de la Fédération Francaise de Judo*[23]  2004
Fig 52-54 one of the most complete works on Seoi biomechanics.

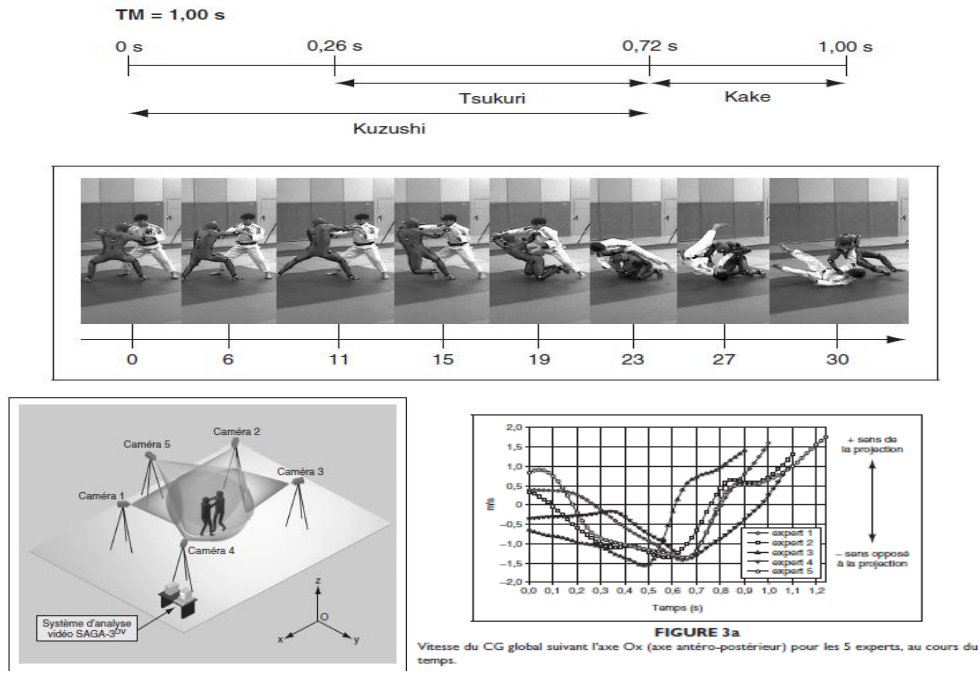

*Fig.52.54 Research layout, motion capture and some results on Suwari Seoi [23]*

Blais, Trilles, Lacoture : *Détermination des forces de traction lors de l'exécution de Morote Seoï Nage réalisée par 2 experts avec l'ergomètre de Mayeur et un partenaire* [24]  2007  Fig. 55-57

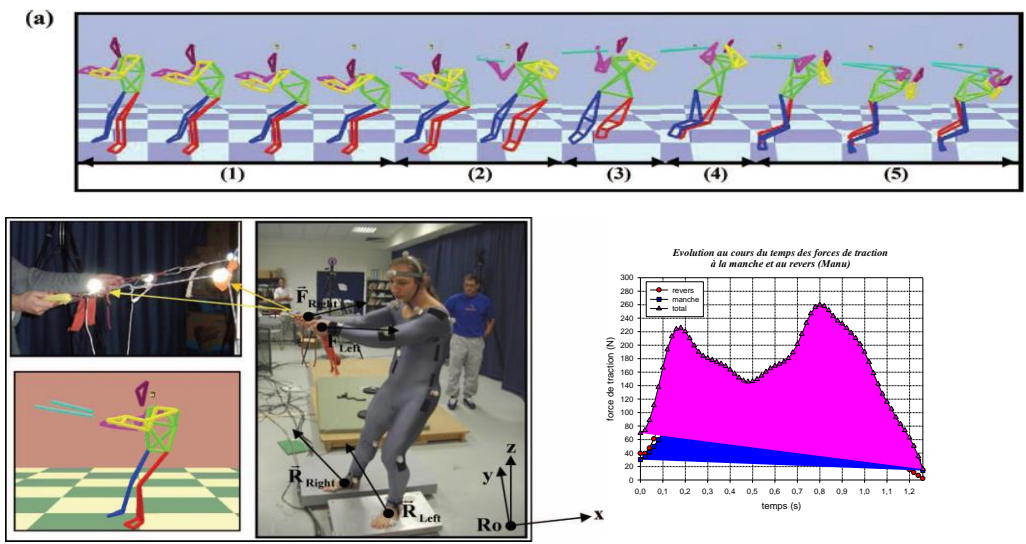

*Fig 55-57 Layout and results of Forces applied in Seoi [24]*

**Germany**

Thiers and Coworkers : *Sport analysis using Shimmer ™ sensors* [25]   2012 Fig. 58

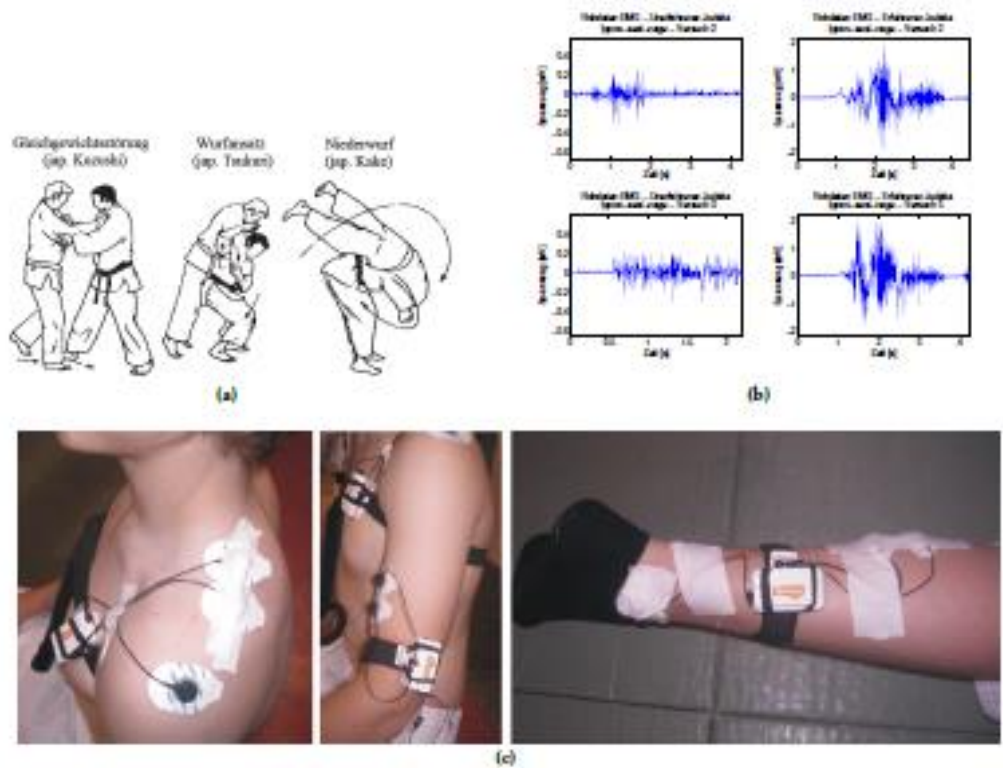

*Fig.58  Research Layout  and some wireless sensor results[25]*

**Italy**

Sacripanti and coworkers : *Infrared Thermography-Calorimetric Quantitation of Energy Expenditure in Biomechanically Different Types of Jūdō Throwing Techniques. A Pilot Study* [26] 2015   Fig.59-62

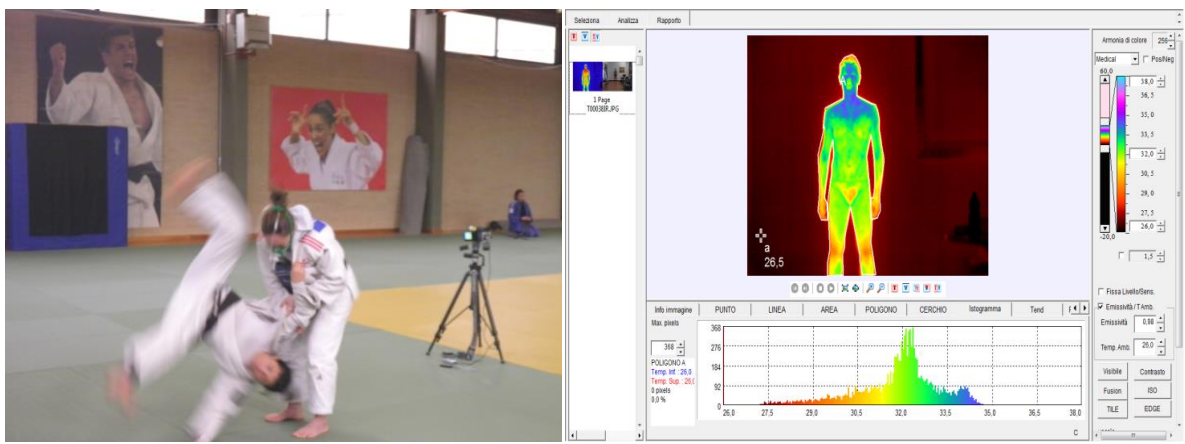

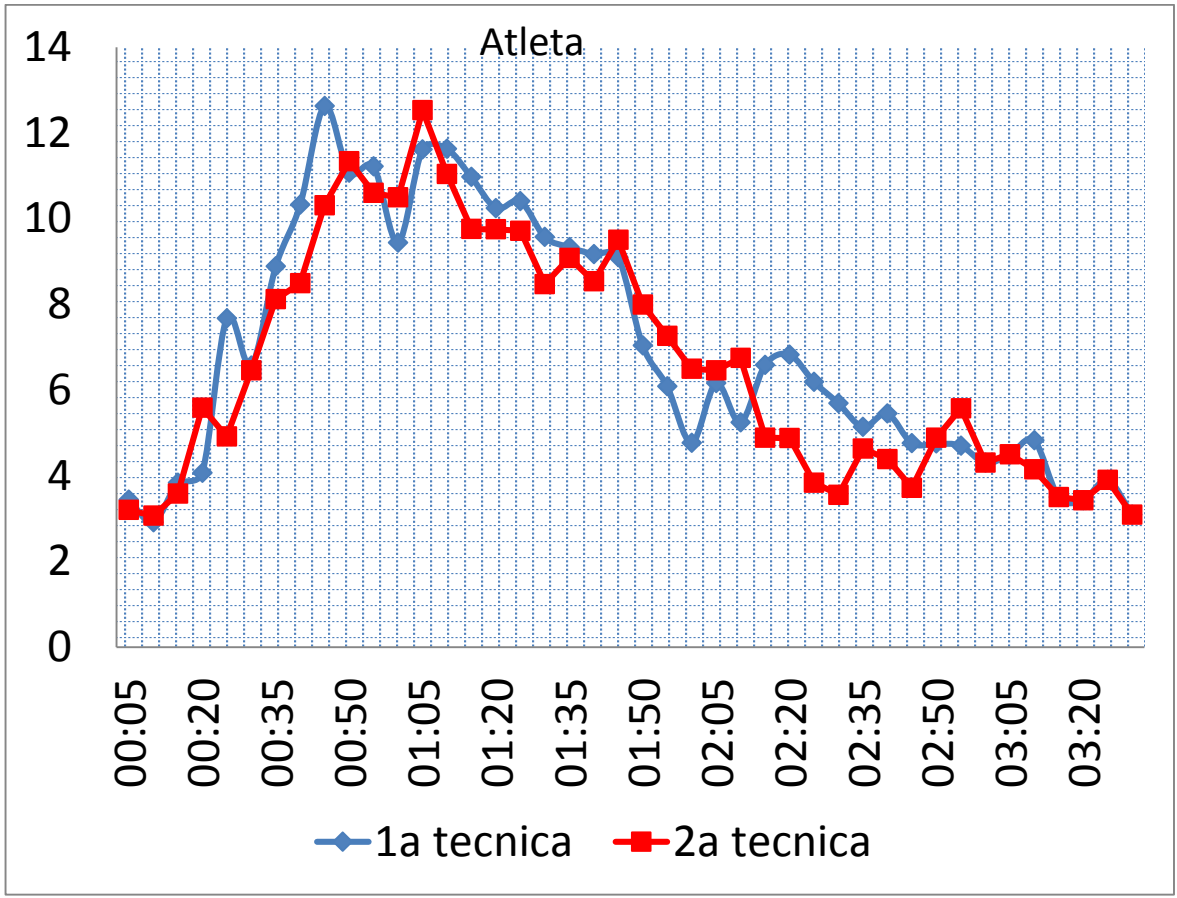

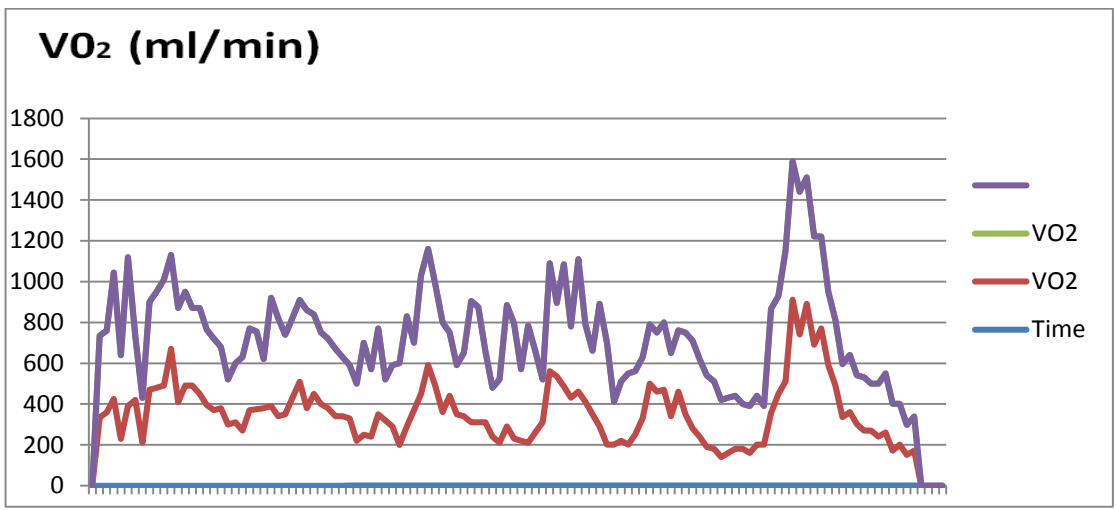

*Fig 59-62 Results of two Metabolimeters and thermocamera for Seoi and Uchi Mata [26]*

**Japan**
Aoki and Coworkers: *Biomechanical analysis of Seoi Nage in judo throwing techniques:*[27] 1986 Fig. 63-66

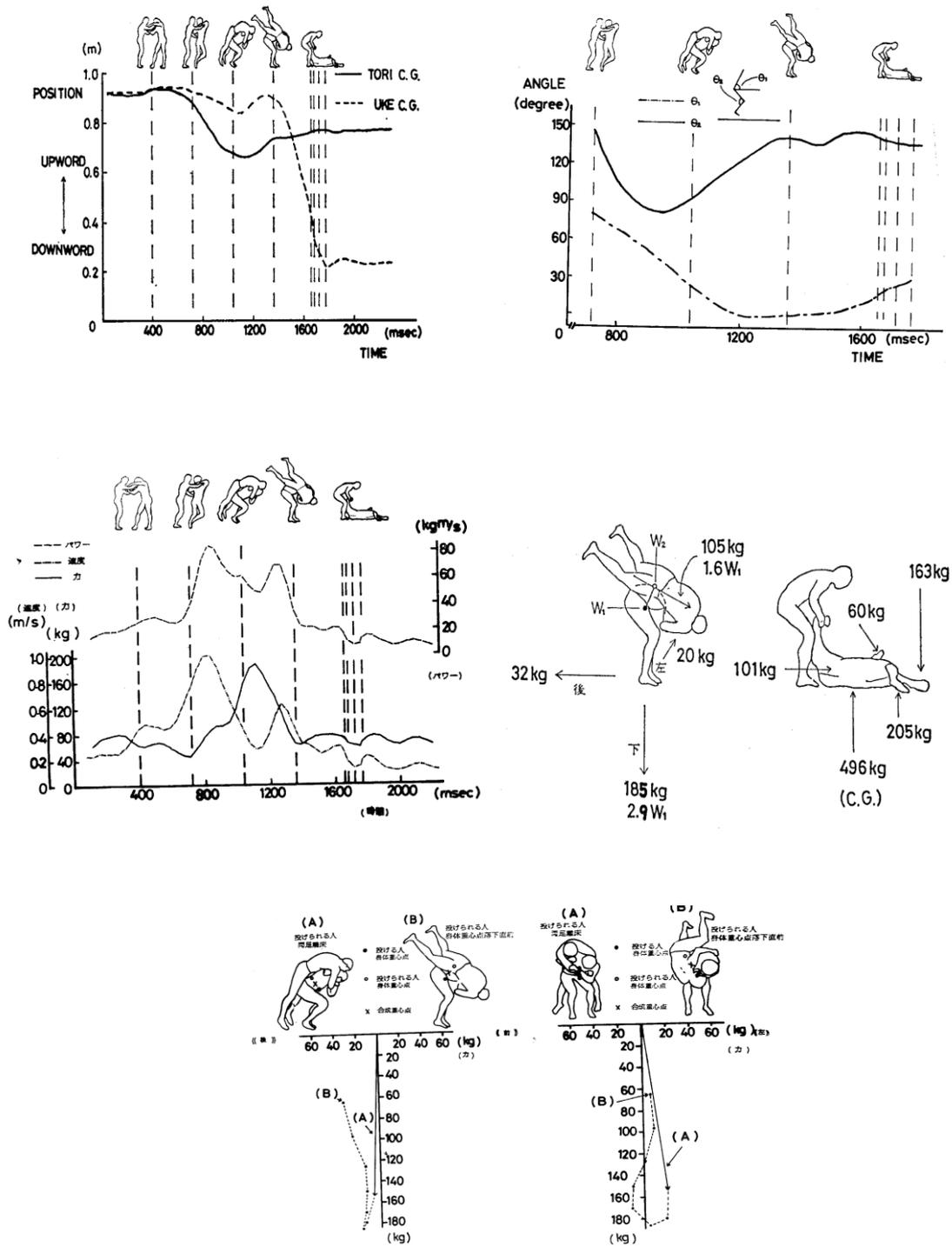

*Fig 63-66 CoM, Angles, Velocity, Power, GRF, for Seoi Nage [27]*

Ishii and Ae : *Biomechanical factor of effective Seoi Nage in Judo* [28] 2014 Fig 67-68

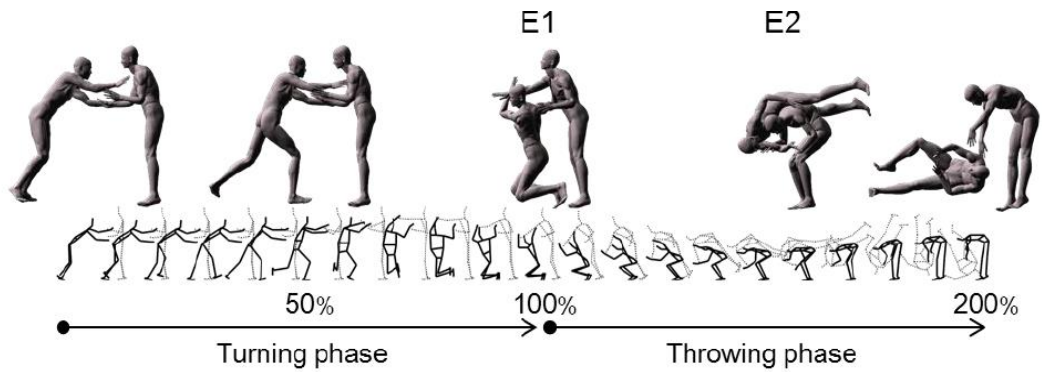

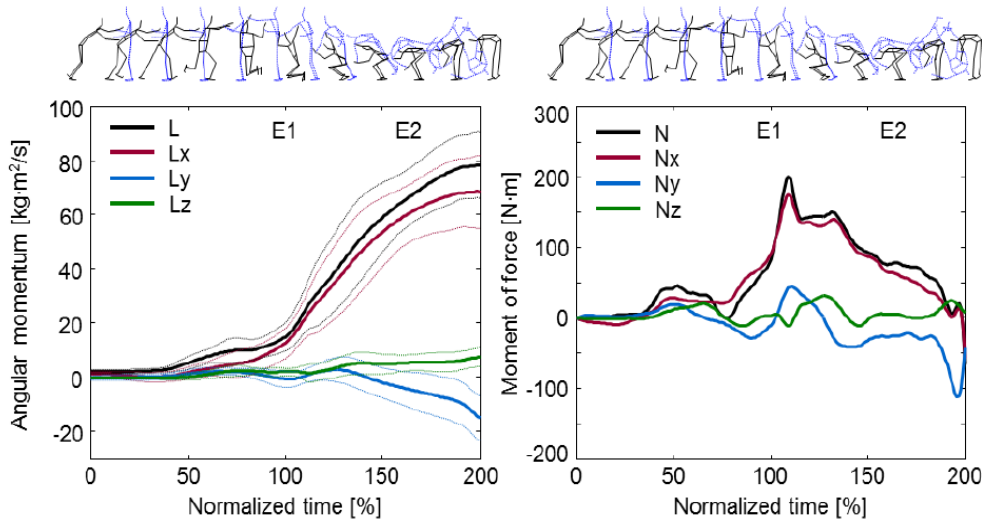

*Fig.67-68 Phases, Angles Velocity and momentum of effective Seoi Nage [28]*

Ishii and Coworkers: *Front turn movement in Seoi Nage of elite Judo Athletes* [29] 2012 Fig 69-70

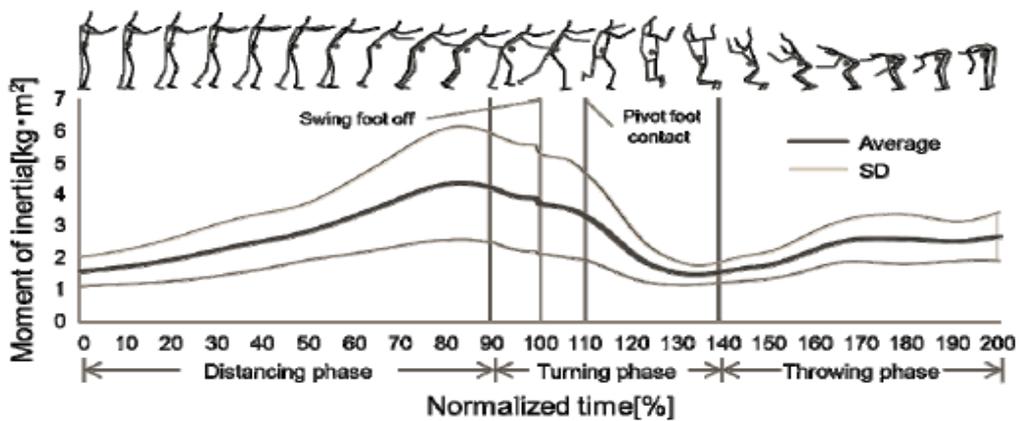

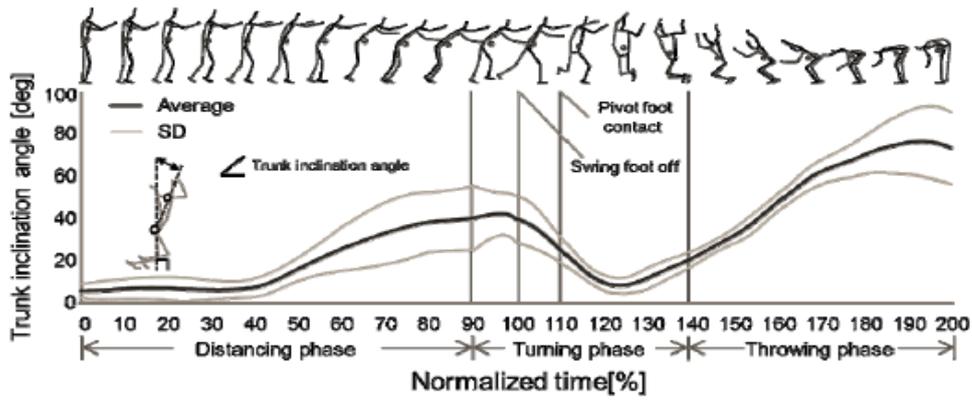

*Fig. 69-70 Momentum and Trunk inclination in Standing Seoi Nage [28]*

**Korea**

Ji Tae & Seong- Gyu : *A kinematic analysis of Morote Seoi Nage according to the Kumi Kata types in Judo* [30] 2006 Fig. 71-72

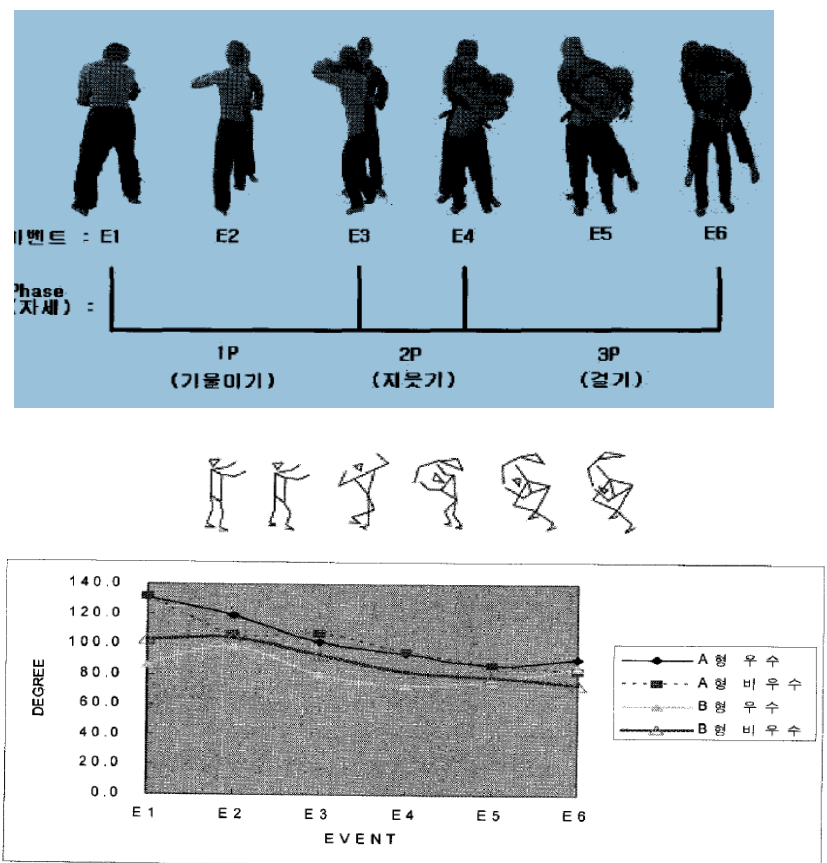

*Fig. 71 -72 Layout and Attack Anglein Morote Seoi Nage [30]*

**Poland**

Chwala, Ambrozy, Sterkowicz : *Tridimensional analysis of the jujitsu competitors motion during the performance of the Ippon Seoi Nage throw* [31] 2013 Fig 73-74

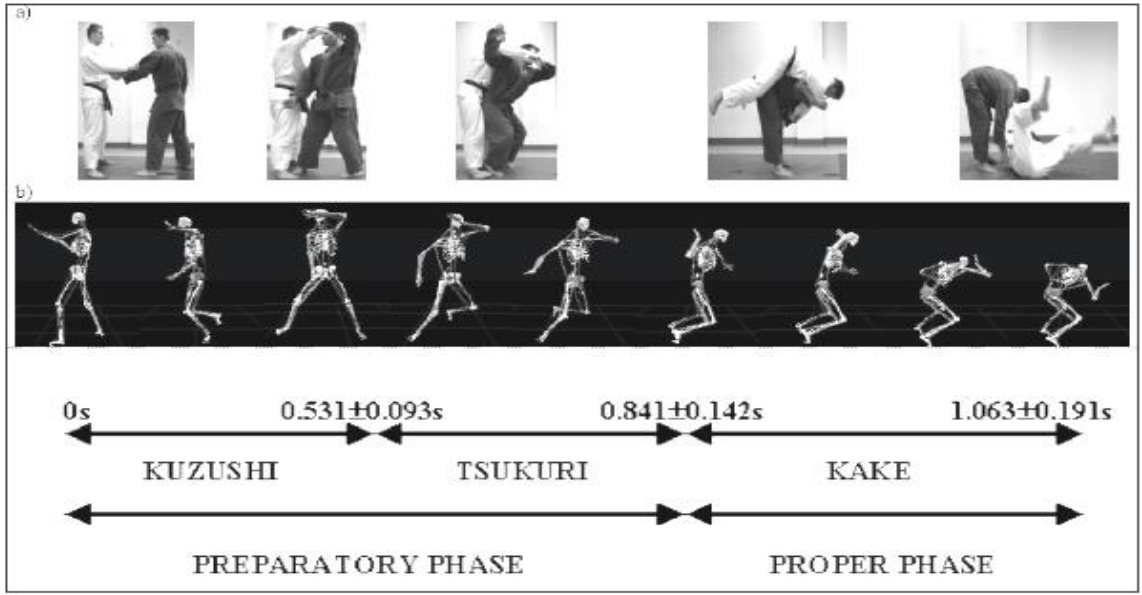

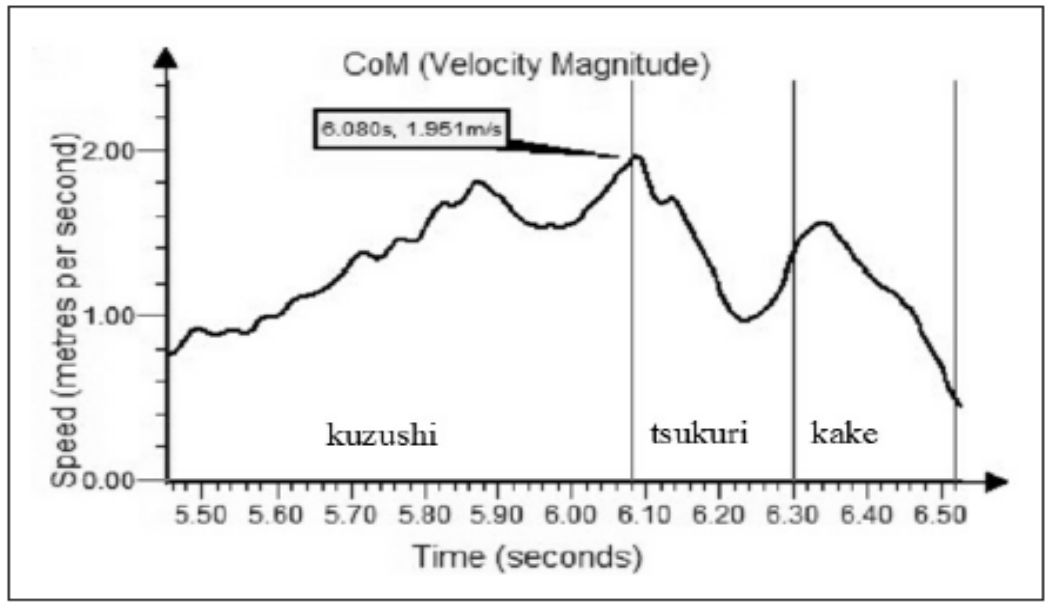

*Fig 73-74 Layout and CoM velocity [31]*

**Portugal**

Peixoto & Monteiro: *Structural Analysis And Energetic Comparison Between Two Throwing Tecniques Uchi-Mata Vs Ippon-Seoi-Nage.* [ 32] 2012 Fig 75-77

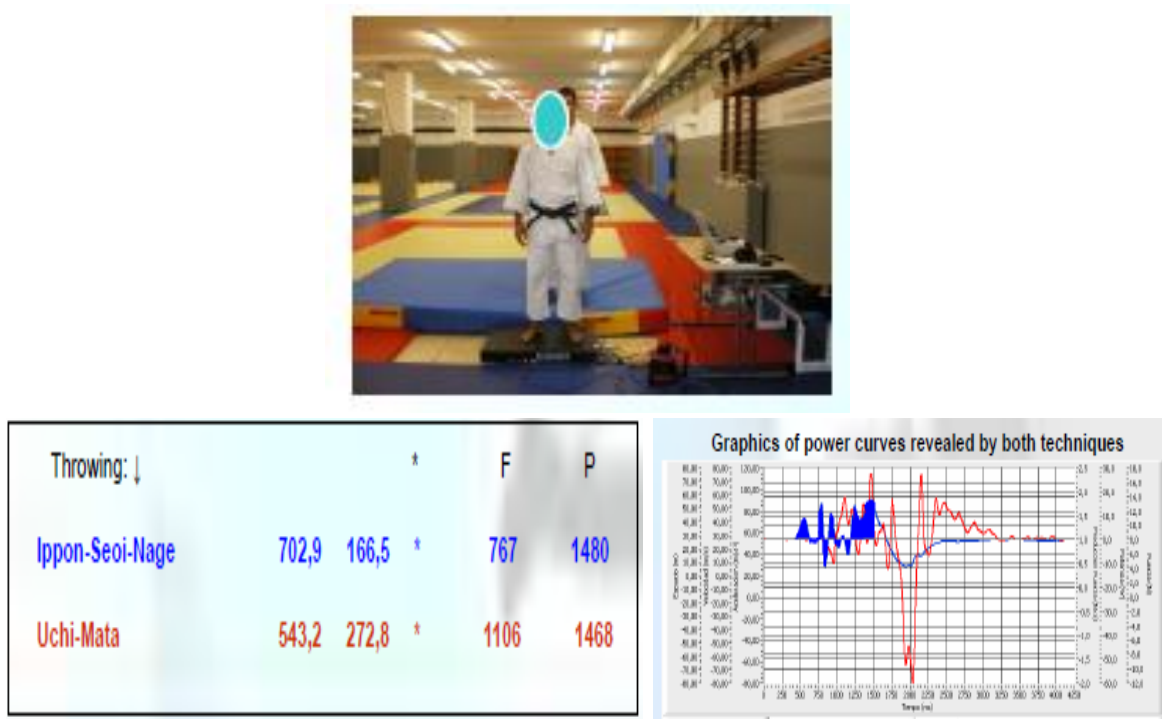

*Fig75-77 Layout and difference in power between Seoi Nage and Uchi Mata [32]*

**Spain**

Gutierrez- Santiago & Coworkers: *Sequence of error in the Judo Throw Morote Seoi Nage and their relationship to the learning process.* [33] 2011 Fig 78

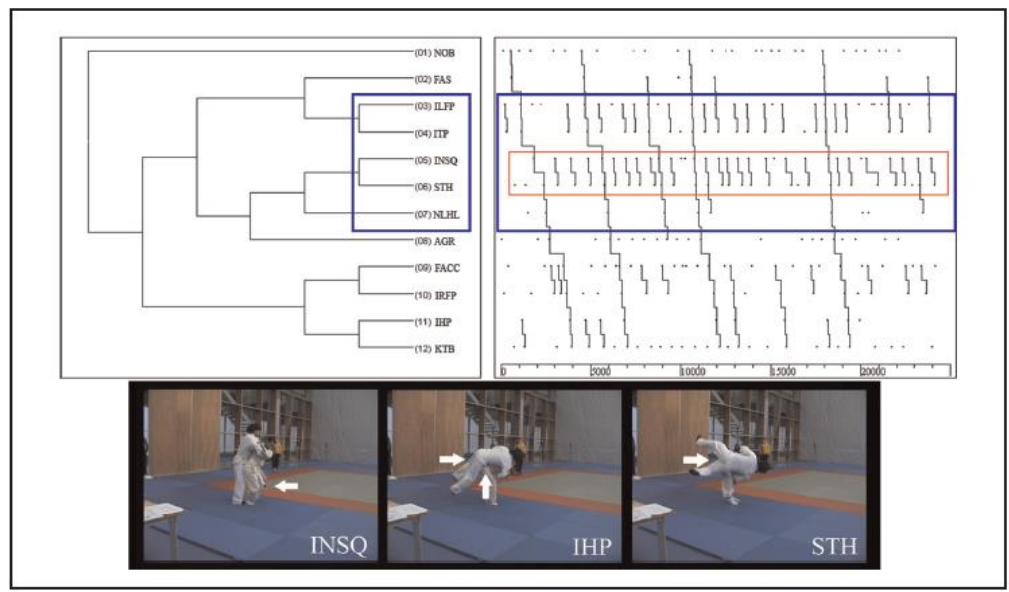

*Fig 78 Analysis of Errors in Morote Seoi [33]*

Carretero & Lopez Elvira : *Impacto producido por la tecnica Seoi Otoshi . Relacion con anos de practica y grado de judo* [34]  2014  Fig 79

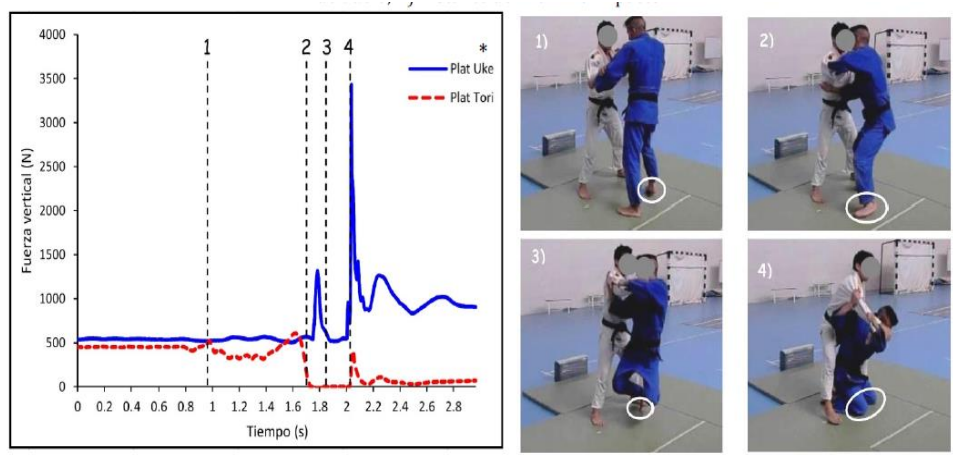

*Fig 79  Knee Impact in Suwari Seoi [34]*

**USA**

Imamura and Coworkers: *A three-dimensional analysis of the center of mass for three different judo throwing techniques.* [35]  2006  Fig 80-81

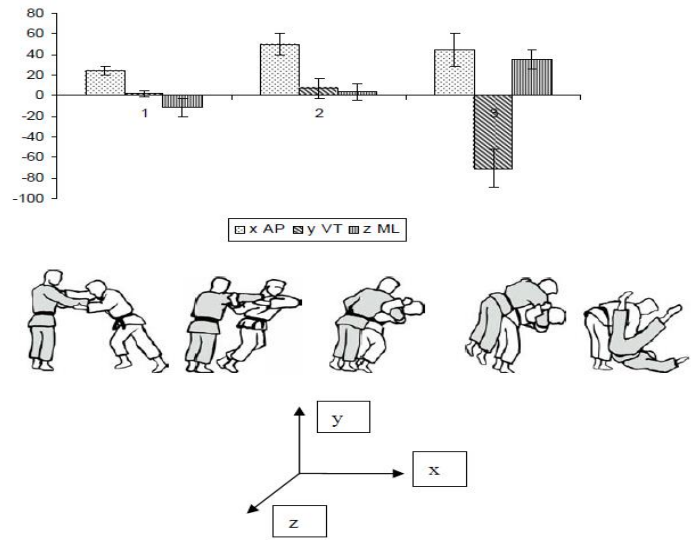

*Fig.80-81  CoM variation and Impulse of Seoi Nage [35]*

## 5. Seoi Direct Attack: Complementary Tactical Tools.

Considering well known the mechanics of throwing by Seoi in static or classical teaching condition, The problems that can arise in real dynamic competition are numerous: at first the difficulty to overcome the defensive strength of arms arranged in Kumi Kata (grips), then because more often the Kano kuzushi (unbalance) concept is not always applicable, the thrower comes face to face with the adversary's break of symmetry, then needs to apply the inward rotation with perfect timing, followed by a collision with the right angle connected with the right line of throwing force applied.

These entire, not only as quality needs a very high coordinative capability, but is applied against an adversary that not agrees with these attempts.

To solve this situation not easy solvable more often the application of throw is partially perfect and athletes utilize some complementary tools to simplify or to refine the outcome of Seoi.

These complementary tools will be subdivided for Standing Seoi, Kneeling Seoi, or for sly applications like Totally Rotational Seoi and Inverse Seoi.

At first as simple notation we remark that for Biomechanics. Standing Seoi, Seoi Otoshi and Suwari Seoi , are not three different techniques but the application of the Lever principle changing only the length of the arm.

## 5.1 Standing Seoi complementary tools in direct attack

*Acting against gravity (lift) detaching feet from the mat + Mawarikomi*
*Makikomi Prosecution*
*Lever applied with Couple enhancement*
*When detached rotary application*

The first tool was among other habitually utilized by the astonishing Japanese expert of Seoi throw during eighty Toshihigo Koga. The following figures show the basic Koga method Fig. 82-85

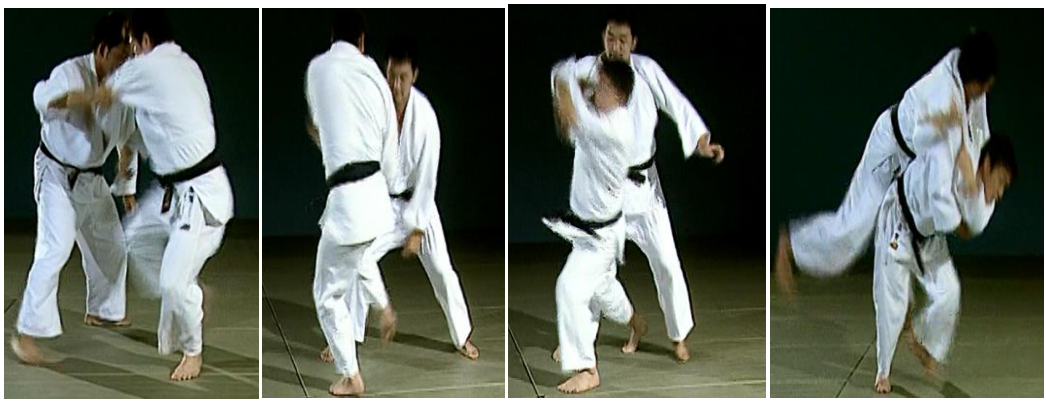

*Fig82-85 Basic Application of the lift tool by Koga*

*Koga System*

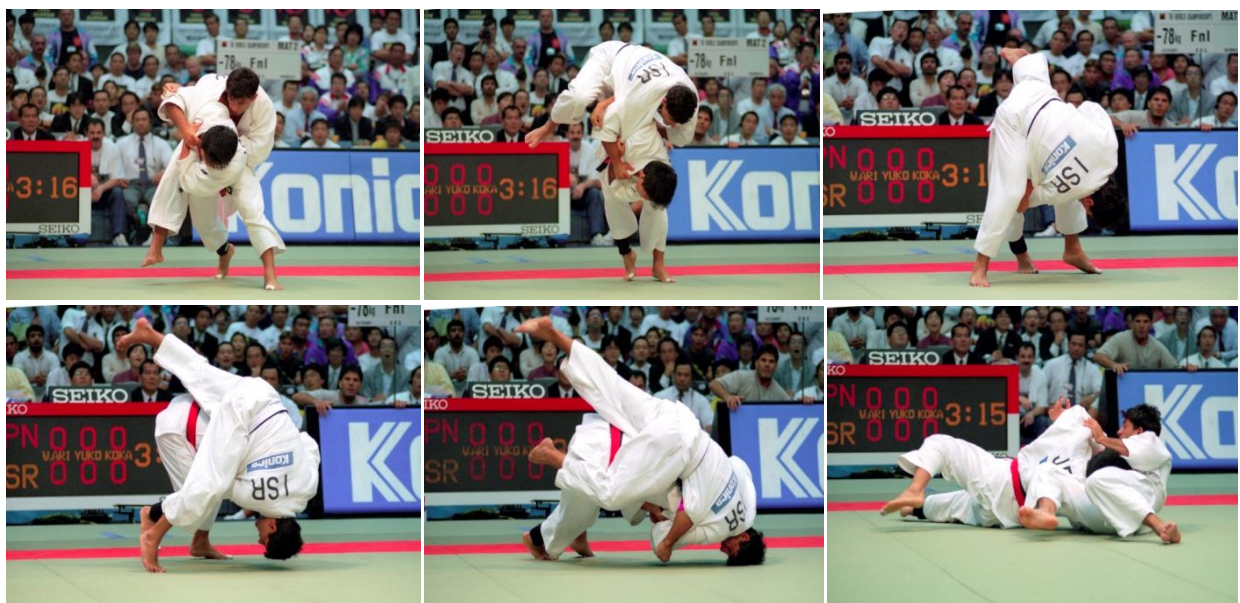

*Fig.86-91 Ippon Seoi with lift Koga System*

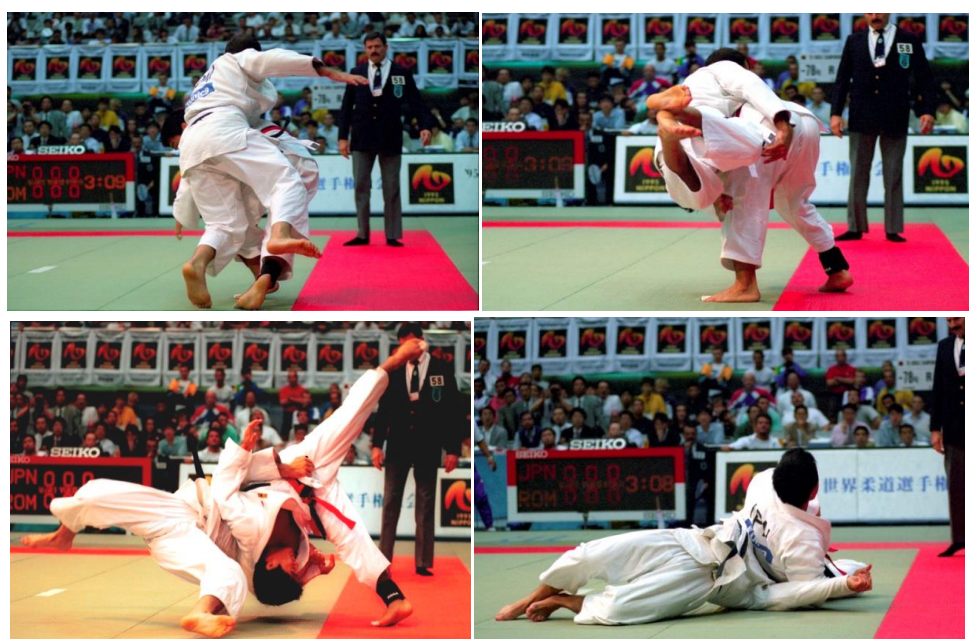

*Fig.92-95 Seoi + Mawarikomi with lift Koga System*

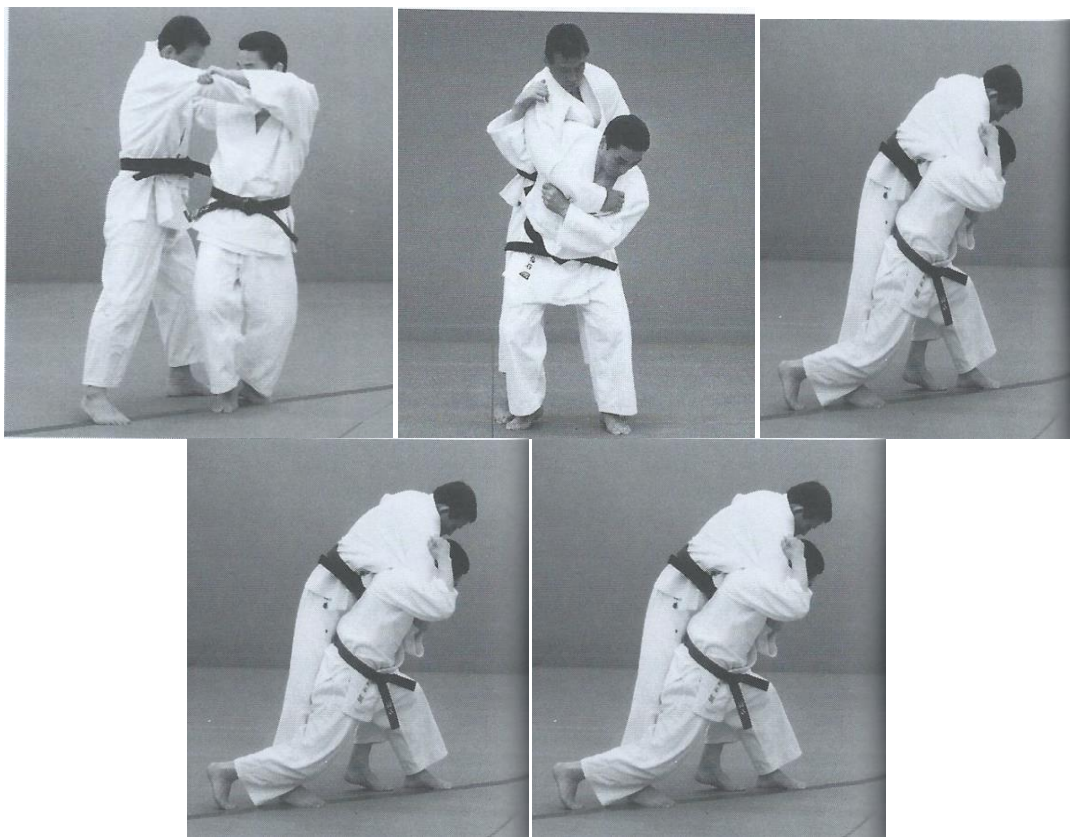

*Fig.96-100 Standing Seoi with Makikomi Prosecution*

*From Ippon Seoi to Uchi Makikomi*

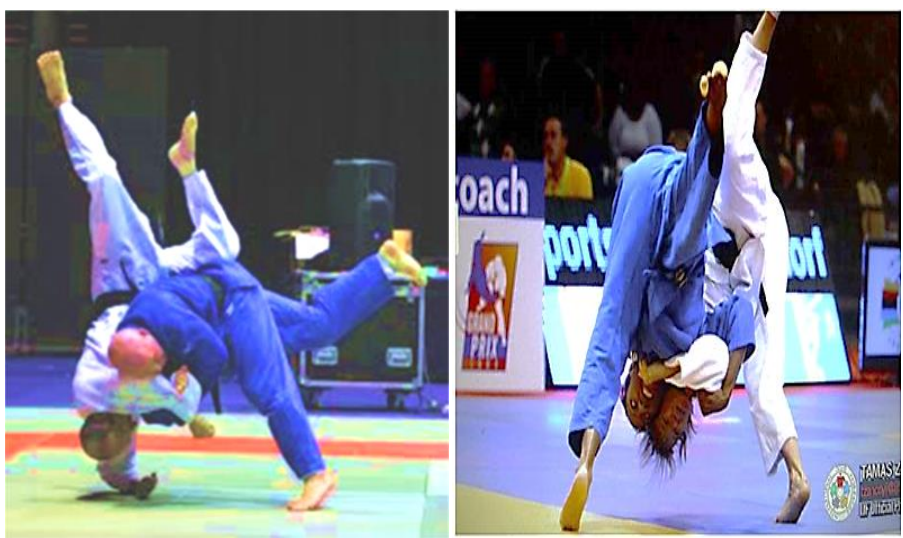

*Fig 101-102 Seoi enhanced by Couple application*

## 5.2 Kneeling Seoi enhancement in direct attack

On kneeling Seoi like Seoi Otoshi , one of the most efficient system to increase the throw effectiveness is to apply throwing forces with a special trajectory: starting in oblique direction and closing with half circle, helped by a body's twisting torque.
In such way the defensive capability of adversary will drop down faster.
This trick was also often utilized by another renowned Seoi Expert Sozo Fuji .

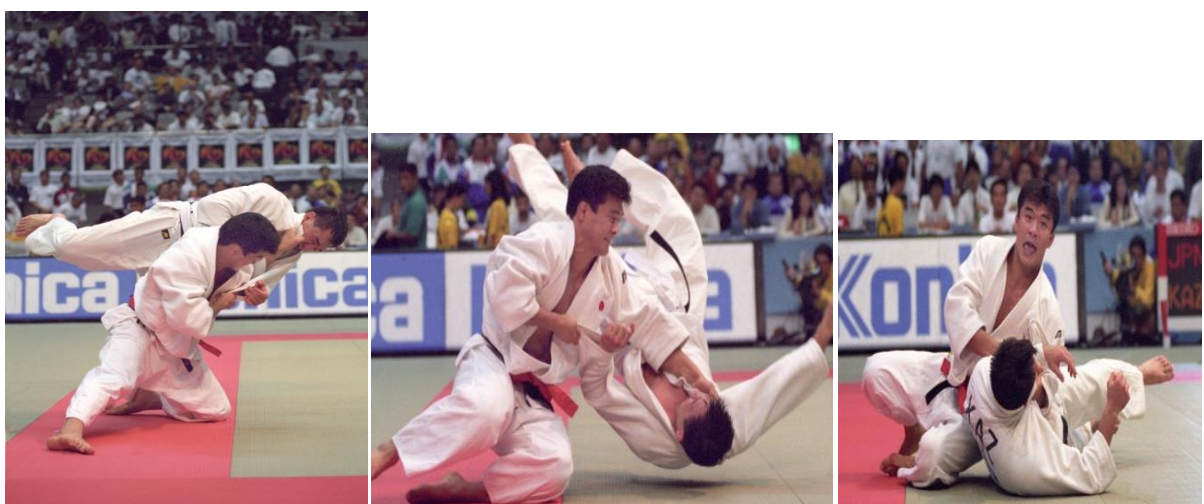
*Fig.103-105  Hidari Seoi Otoshi Koga in Fuji style*

## 5.3 Tools to enhance effectiveness of Suwari Seoi in direct attack

*Acting against gravity (lifting) detaching feet from the mat to avoid any possible defense*
*Acting against gravity (lifting torque) refining throwing action to obtain Ippon*
*Twisting Torque Application*
*Crosswise Application*

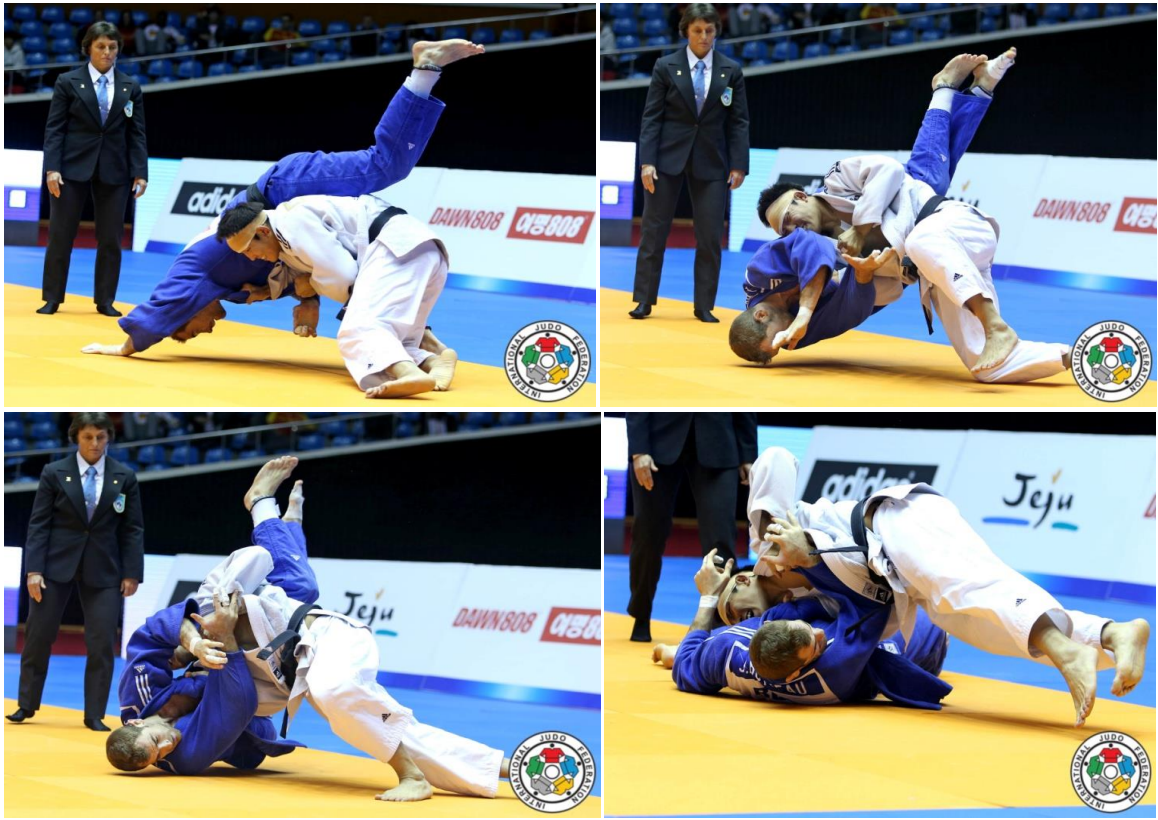

*Fig.106-109 Suwari Seoi with lift up torque to refine the final action*

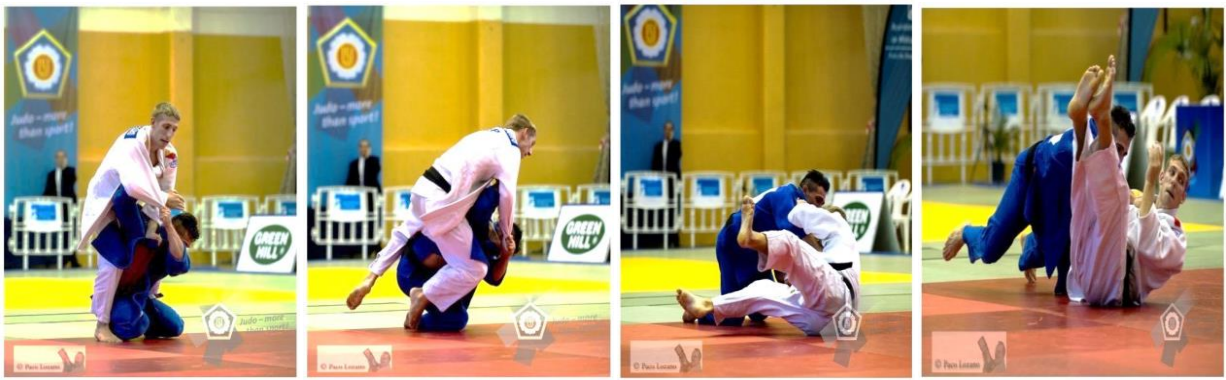

*Fig110-113 Suwari Seoi enhanced by an horizontal twisting torque application*

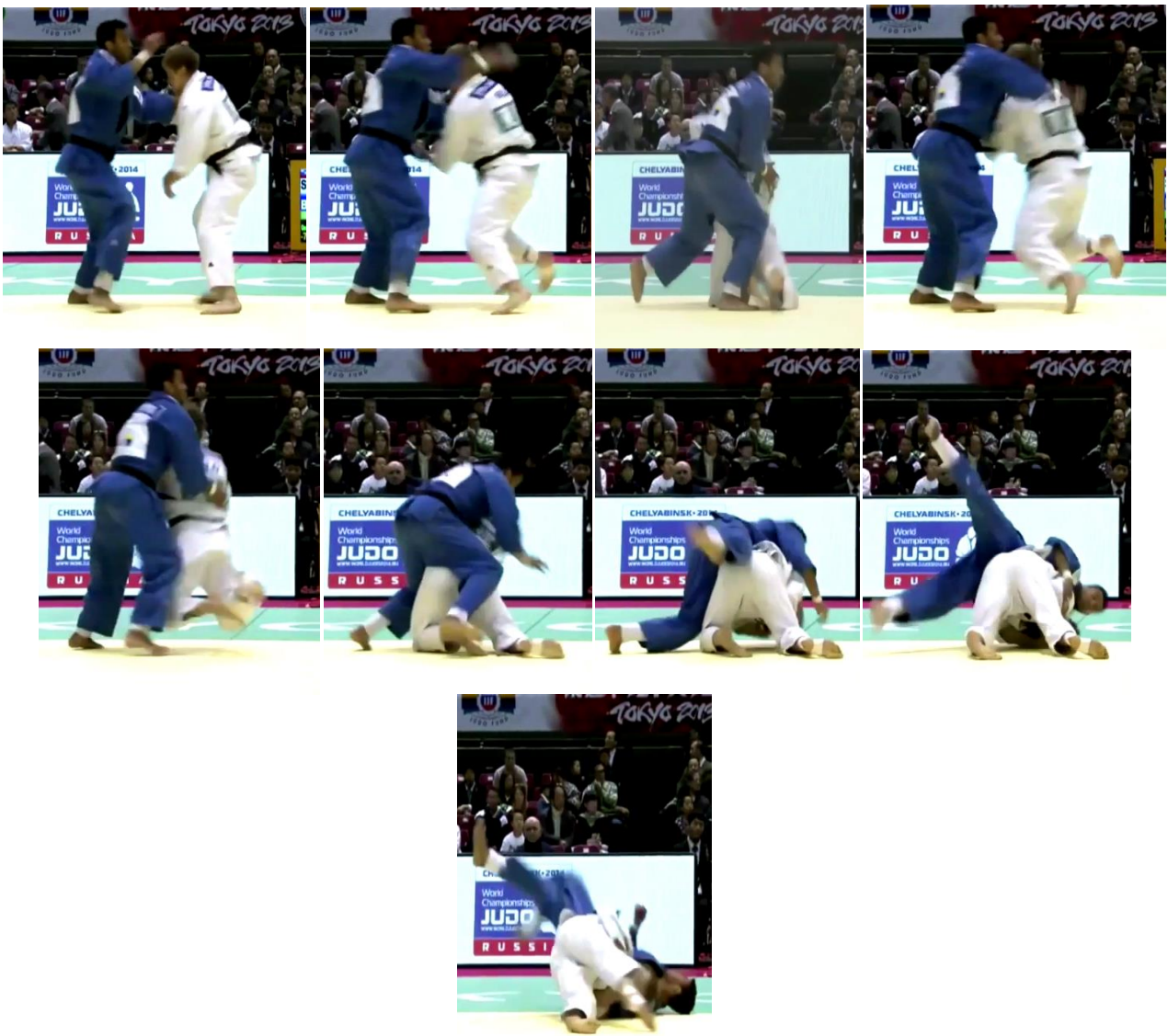

*Fig 114-122 Enhancement of Suwari Seoi by a crosswise application*

But way athletes need to apply such complementary tools to enhance technique effectiveness? Normally during high dynamic competitions situations, applying Suwari Seoi, is difficult to manage easily the throwing force final direction, for two reasons:
A) Dynamicity of a throwing action
B) High defensive capabilities of athletes
These two are the main reasons to apply tactical support tools to refine action in competition, because during the fall down Tori have few control of Uke movement, this is way the end is a situation of multiple direction choice for Tori and this is also the main difficulty to manage this throw as shown in the next figure 123.

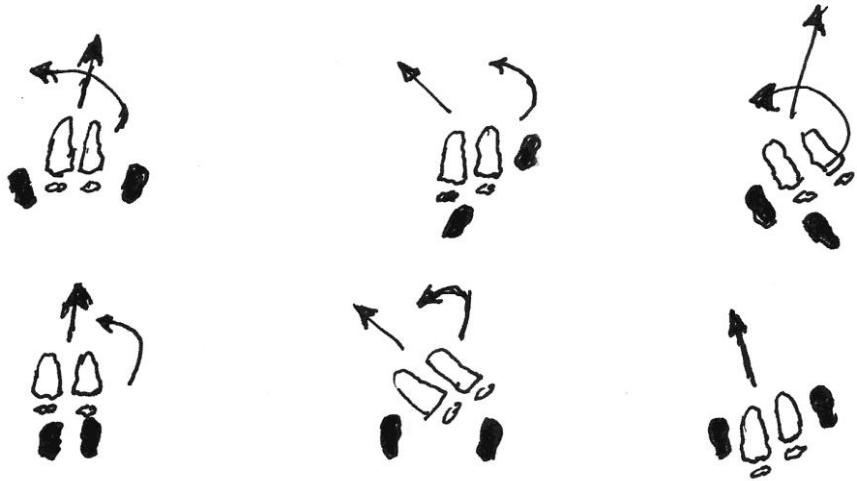

*Fig123 footprints (Tori legs/ Uke's feet) showing the multiple direction choice of throwing force*

## 6. New ways: Reverse and Rotational Application

In the last years the research of new way to enhance the effectiveness of judo throws developed also in Japan interesting non orthodox solution of competitive situations.

A worldwide well known new application is the so called Reverse Seoi, that seems the evolution of Ude Gaeshi presented both in Kudo *Dynamic Judo* and Sato and Okano *Vital Judo* [15] with some light differences. Fig.124

These new application need a careful acrobatic training and a precise timing for application.

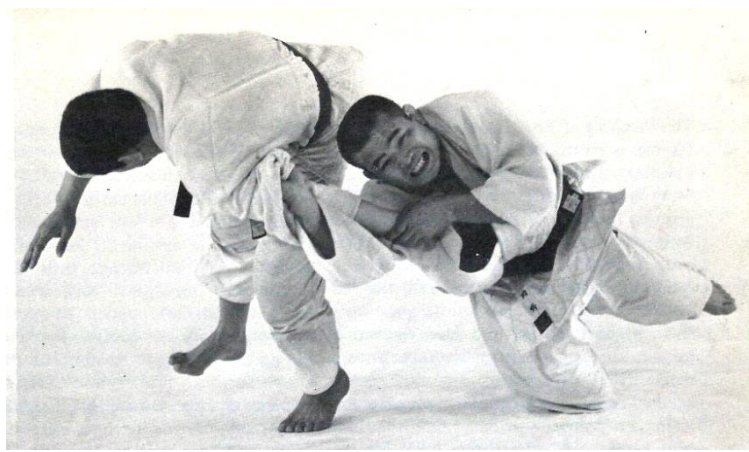

*Fig 124 Ude Gaeshi*

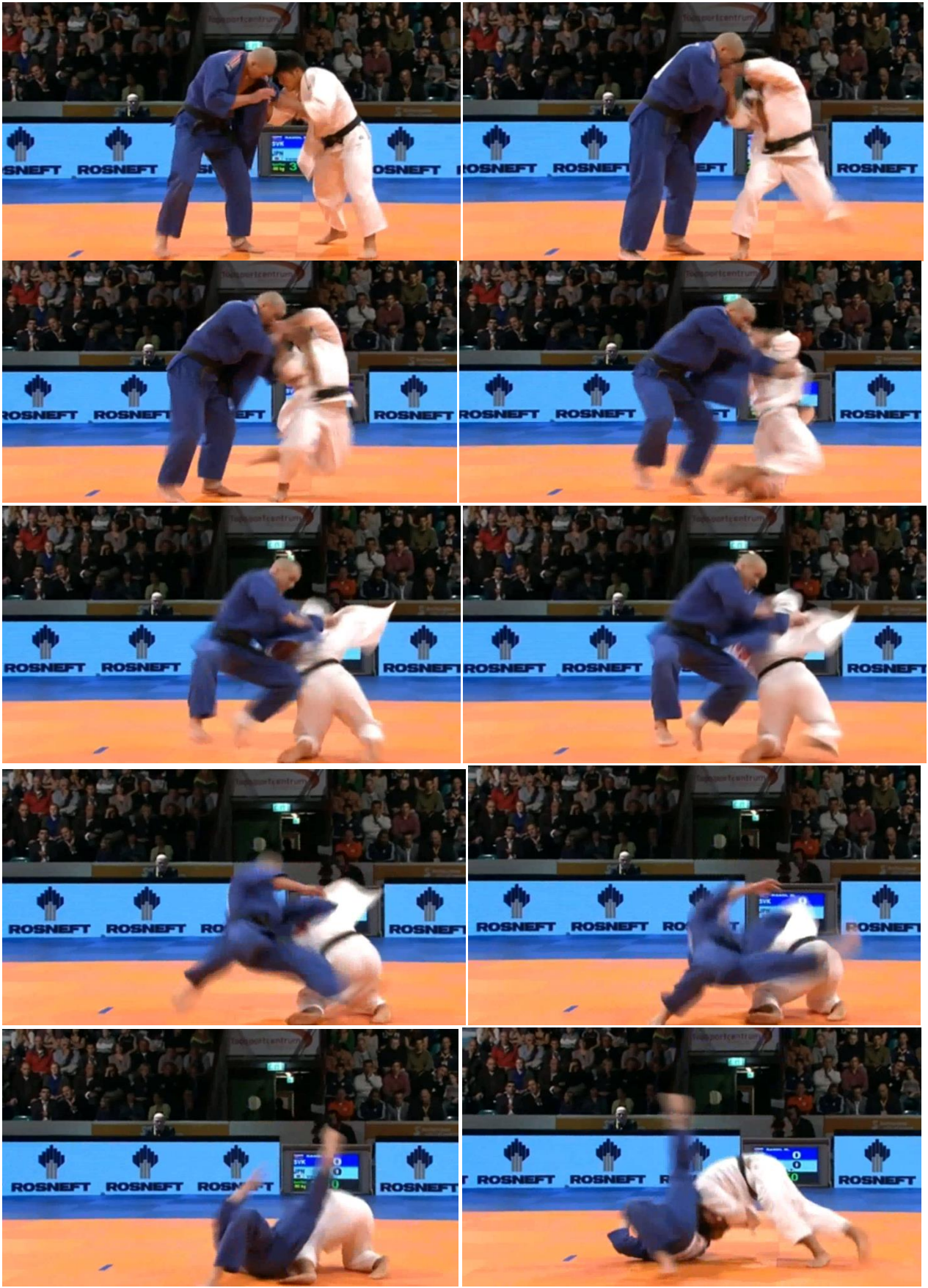

*Fig 125-134   Reverse Seoi  applied with a 180° of rotation*

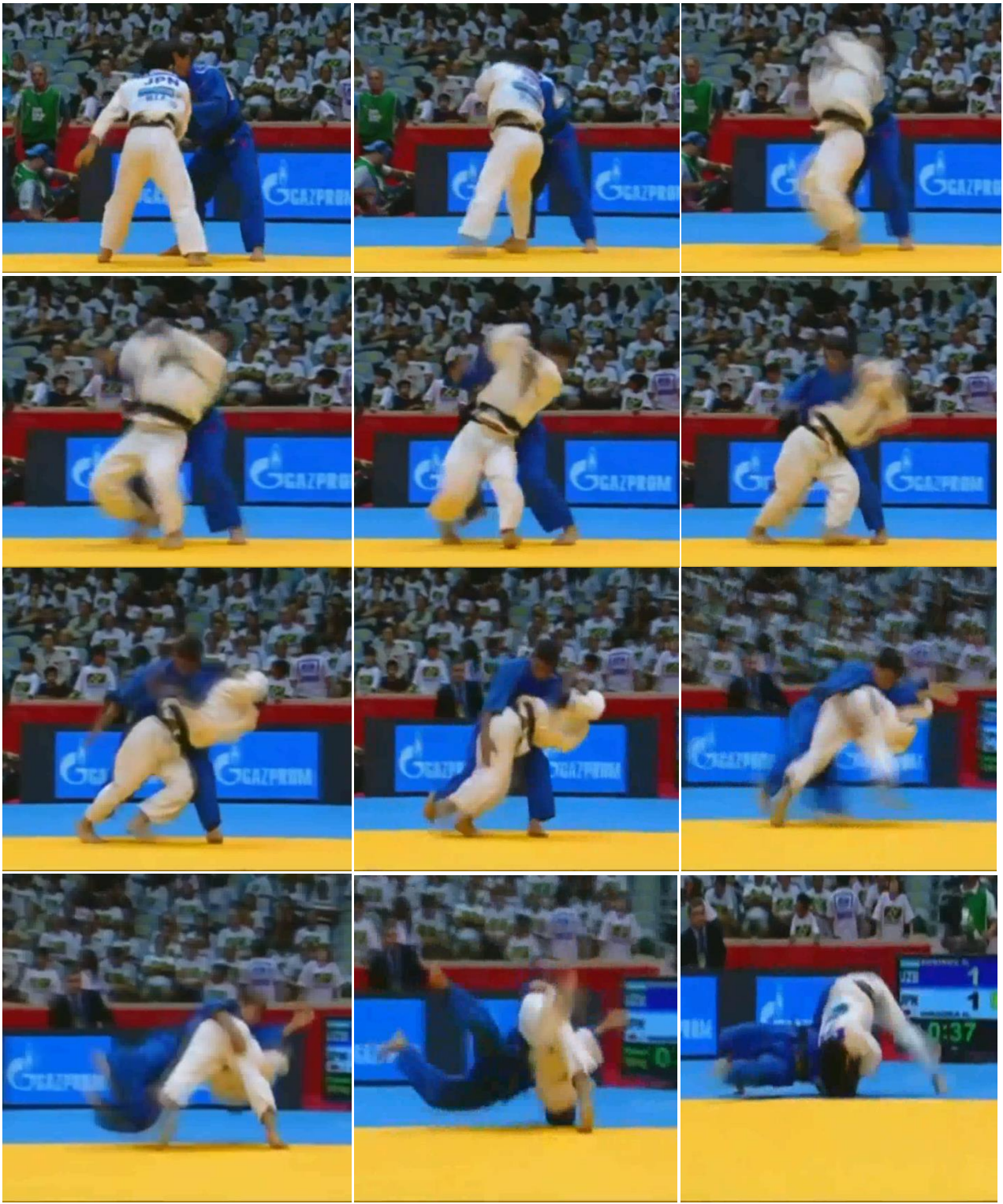

*Fig 135-146 Reverse Seoi with 360° rotation by pivot on the head*

## *Standing Spinning Variation*

This not usual kind of Seoi is very effective and at light of his mechanical properties it is possible to perform it without Kuzushi.
The application of Rotational Dynamics changes the basic Biomechanics of this Seoi throws.
 It is a special form of standing Seoi always a lever technique, but not with minimum arm, although with maximum arm.
This means not high expensive but with minimum metabolic energy.
This very unusual Seoi is very tricky and very effective however it is easy to stop it, but in one only way that is for inverse very easy to change the attack, following with Ko Uchi or O Uchi.

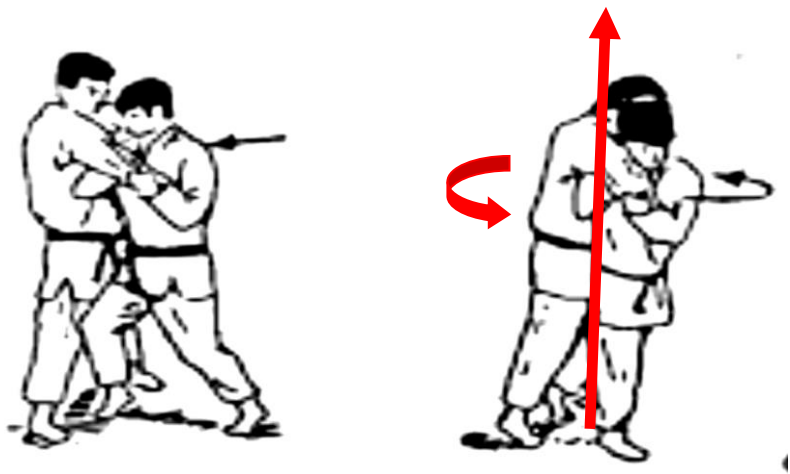

*Fig.147 Spinning Seoi*

# 7. Seoi: Physical and Biomechanical framework

At first the unifying biomechanical vision shows us that the three techniques (in Japanese classical vision) utilized in competition are always the same technique in which is only changed the arm length with the goal ( among others) to decrease energy consumption.
 The basic principles of Lever, is easy to find Fig.148, but analysis of Seoi throwing actions is complex matter. Easy again is to understand the best effectiveness for the throw, in term of angles between hips hitting actions and throwing force direction, because Tori apply a momentum, hitting direction and throwing force direction must be parallel.
However Seoi mechanics shows aspects not totally well known, like: almost-plastic collision between extended soft bodies and dynamics of bodies with variable rotational inertia. Usually Kano's Kuzushi concept is inapplicable in Competition. In classical terms rotational *Kuzushi-Tsukuri* is the practical evolution and presupposes a condition of highly dynamic motion. Competitive Judo recently uses these rotational concepts as the most natural and appropriate tools to minimize both efforts and energy.
However for Seoi, from the Biomechanical point of view, in competition counts the most intriguing concept of Breaking Symmetry [36] followed by inward rotation and collision.

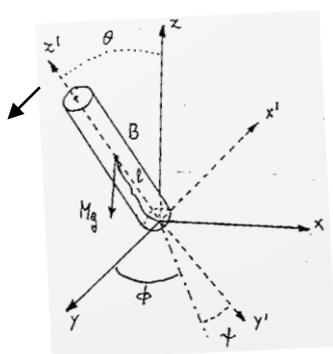 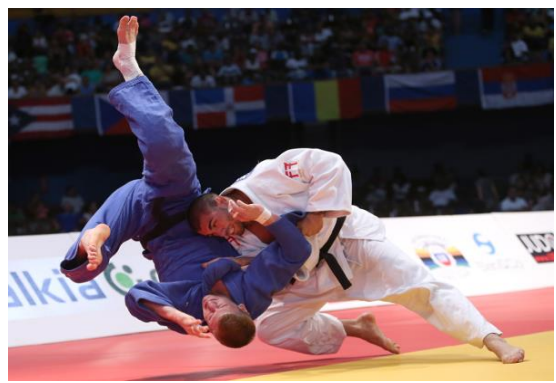

$$\tau = F \wedge r$$

*Fig 148 Mechanical model of Seoi*

*Interaction*
As already express in other papers, [36] the Competition of Judo can be analyzed with the standard physical method, by means of two wide areas One Couple of Athletes motion ( shifting paths analysis) , and Interaction between Athletes.
The interaction study will be bounded at the application of Seoi only, no taking in consideration: other throws, connection from Standing and laying position, groundwork techniques, and so on. ,

*Grips*
These are the main system utilized by Tori to contact Uke and shorten distance between athletes, and the main system utilized by Uke to stop this action.
Biomechanically speaking arms have three weak points their joints: wrist, elbow and shoulder.
 As general rule, arms are less able to oppose to up and down movement, than push/pull actions.
Usual trick to overcome strong grips is: *to apply a downward pressure followed by a sudden upward impulse with a simultaneous sliding of the body in inward rotation under grips.*

*Lift*
Lifting action aims at decrease friction between feet and mat, without contact (friction) Uke cannot apply any defense. The only defensive possibility is to grips the Tori's body or judogi to avoid the projection. This gripping action generally is overcome by a couple actions (Mawarikomi). As analyzed for Suwari Seoi lift have a refining function.

*Inward Rotation*
After the Break of Symmetry, Seoi needs inward rotation. Tori taking firmly a contact point performs a fast inward rotation and collides applying the tool (lever) to throw Uke.
All trajectories are similar. There are not closed form solutions to the Equation of motion.
The two Bodies that play judo are mainly connected by arms that apply push pull forces in every direction. In this sense we can compare the body drifting apart or getting close at a situation of dynamic of a body in a field of elastic internal forces.
. The Bertrand theorem assure us that the only closed orbit for the elastic generalized force
$$F = -kx^\alpha \quad (3)$$
There are only for α = 1; or α= -2 as already demonstrate in [37].
How it easy to see in the next figure all trajectories are similar to it, because Tori taking firmly a contact point performs a fast inward rotation to apply the lever and throw Uke.
It is interesting that always speaking in term of elastic field the class of trajectories applied by Tori are similar to a capture trajectory of a particle by another in elastic field with internal forces
$$F_{ab} = -F_{ba} = F \quad (4)$$
In this case the two particles act like the two athletes bodies during the inward rotation throw. [38]

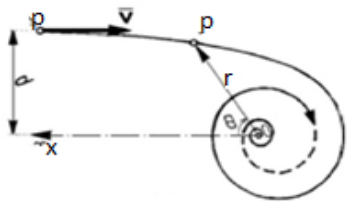

*Fig .149 Trajectory of a capture particle similar to Tori's throwing inward trajectory in judo interaction* [38]

*Quasi-Plastic collision of extended (soft) bodies.*
Most often the end of the previous trajectory flows into a projection by a lever techniques that starts with a collision, which can be considered a quasi plastic collision because the two athletes are strictly connected together, but obviously their bodies are not merged, then it is a collision quasi-plastic of two extended bodies. In this case considering both athletes of more or less equal mass, the equations are very easy to obtain. [39]
If they have different starting velocities, after the contact they move connected till to the fall down that can be considered as a free fall.
In this case considering always negligible the gravity force, that increases greater and greater his importance into the motion after the collision, till to landing of Uke body (free fall); it is possible to write for the early instants of the collision, remembering that is a rotational impact: [40]

$$mv_1 + mv_2 = 2mv \qquad (5)$$

the conservation of angular moment give us $I_1\omega_1 = (I_1 + I_2)\omega_f \qquad (6)$

or $\dfrac{\omega_f}{\omega_1} = \dfrac{I_1}{(I_1 + I_2)} \qquad (7)$

the impact is totally inelastic, and the loss of kinetic Energy is:

$$\Delta K = \frac{1}{2}I_1\omega_1^2 - \frac{1}{2}(I_1 + I_2)\omega_f^2 = \frac{1}{2}\frac{I_1 I_2}{I_1 + I_2}\omega_1^2 \qquad (8)$$

### *Full rotation with free fall*
Complex rotational application of judo throwing techniques that will be analyzed in this section are connected to the tactics of direct attack applying a lever techniques with a fast and complete inward rotation, like spinning top at variable mass or vortex

It is, for Tori as observer, a clear study of Forces applied in a rotating reference frame.
At first it is important to evaluate the velocity transformation formula from the inertial to the rotating frame, this means how the speed is evaluated by people (public as observer) and Tori as observer during the execution of a rotating throws, as already demonstrate in a previous paper [41] the result is:

$$v = V + \left(\frac{dr'}{dt}\right)_O = V + (v' + \omega \wedge r') \qquad (9)$$

It is important to evaluate the general equation of motion of this variable mass spinning up, remembering the classical Newton approach to the rotational dynamics, [42] we can write:
The torque on the first athlete $\tau$ will be

$$\tau = r\,F \qquad (10)$$

Where $r$ is the radius between the centre of mass and the point where the force $F$ is applied.
In term of rotational dynamics this equation can be written also as:

$$\tau = I\frac{d\omega}{dt} = mr^2\frac{d\omega}{dt} \qquad (11)$$

It is very easy to solve this equation if the athlete is up-righted like a symmetric spinning top, because as long as the torque is applied in such a way as to increase (or decrease) its rotational speed around the $\hat{z}$-axis, this is just a one-dimensional equation and offers no surprises. [43]

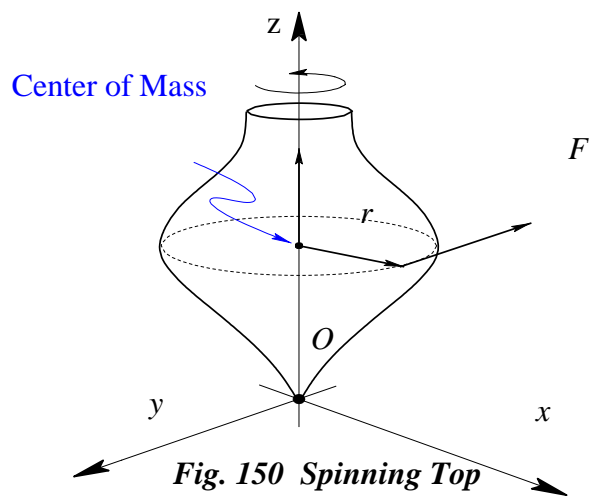

*Fig. 150 Spinning Top*

It is necessary to remember that if we use the Euler representation of a rotating rigid body they produce a non linear system of equation not always resolvable.
However interesting are the equations of motion of the Euler angle found by Garanin [44], which are shown in the next:

$$\dot{\theta} = \left(\frac{1}{I_1} - \frac{1}{I_2}\right) L \sin\theta \sin\psi \cos\psi$$

$$\dot{\varphi} = \left(\frac{\sin^2\psi}{I_1} + \frac{\cos^2\psi}{I_1}\right) L \qquad (12)$$

$$\dot{\psi} = \left(\frac{1}{I_3} - \frac{\sin^2\psi}{I_1} - \frac{\cos^2\psi}{I_1}\right) L \cos\theta$$

Note that equations for $\dot{\theta}$ (nutation) and $\dot{\psi}$ (spin) form an autonomous system of equations that can be solved as first step, after that the equation for the precession $\dot{\varphi}$ can be integrated using the previous found spin, obtaining the demanded solution.

If, however, athlete *tilts* [45] his rotational axis through an angle $\varphi$, as shown in Figure 151, the situation gets a little more complicated and a lot more interesting respect to a spinning top.
In the case of the tilted athlete shown in Figure 152, gravity pulls down on the centre of mass of the athlete, which would pull a non-spinning athlete (because it is in unstable equilibrium) downward and simply increase the tilt angle $\varphi$ as the athlete falls down.
Normally in a spinning top the torque, and thus the change in the angular-momentum vector, is perpendicular to the axis $\hat{u}$, which leads the top to move "sideways" in a circle around the *z*-axis, and this motion is called *precession*, but this well known phenomenon is nullified in our case by the increased mass of the system that after the (quasi-plastic collision) firmly connect together the two athletes bodies.

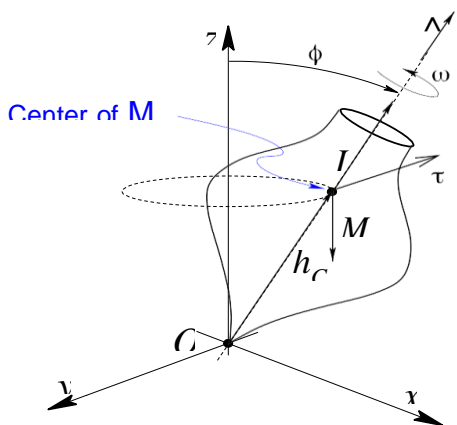
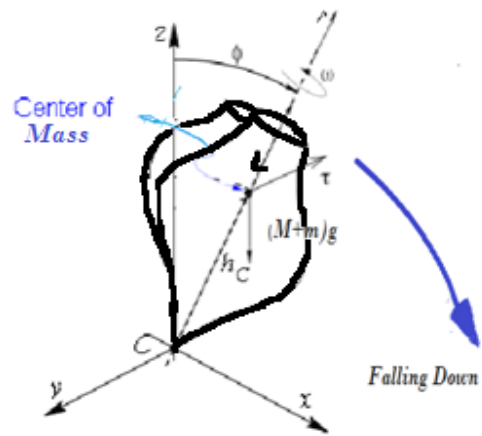

*Fig.151  Tilting Spinning Top;*          *Fig. 152  Variable Mass Falling Spinning Top.*

The three - dimensional equation becomes:
$\tau = r \wedge F \qquad [16]$

$$\tau = \frac{d(I\omega)}{dt} = \frac{d[(M+m)\omega]}{dt} \qquad (17)$$

Abruptly after contact, mass more or less double, like a system at variable mass,[45] velocity drops down and external gravity force, overcoming the potential precession motion, helps the bodies to fall down.
For Lever throws like Seoi tactical tools are lift or rotational applications or hybridization by a supplementary torque
.

## 8. Conclusion

The comparative analysis of Seoi throw made at light of Biomechanics let us able to summarize some remarkable properties useful, for Coaches and Teachers.

Such properties that are shown in the following are defined for each type of throw in the condition of direct attack during competition without any help of complementary tools.
.

### Standing Seoi

1) *Energetically expensive*, *(the arm of the lever is shorter)*.
2) *Most important Kuzushi* *( the Uke stable standing position needs kuzushi)*.
3) *Difficult Tsukuri (it is a % of Tori's body that fits for the Uke's body). %U>%T means a good Kuzushi*
4) *Hard in overcoming Uke's grips* *(best to utilize rotational actions)*
5) *Needs high rotational speed*
6) *Easier to avoid*

### Kneeling Seoi

1) *Energetically less expensive*, *(the arm of the lever is longer)*.
2) *Less important Kuzushi* *( the drop down movement acts as kuzushi)*.
3) *Easier Tsukuri* *(it is Uke's body that fits for the Tori's body). Normally %U>>%T*
4) *Useful in overcoming Uke's grips* *(arms are less able to stop up and down movement, than push/pull actions)*.
5) *Useful in increasing rotational speed*
6) *Most difficult Avoidance vs drop action*

### Reverse Seoi

1) *Energetically favorable*
2) *Less important Kuzushi* *( the drop down movement acts as kuzushi)*.
3) *Null Tsukuri*
4) *Useful in overcoming Uke's grips*
5) *Need high rotational speed*
6) *Most difficult Avoidance vs drop action*
7) *Inverse Trajectory* *(difficult to handle)*.
8) *Applied as psycological trick*

### Spinning Seoi

1) *Energetically very favorable*
2) *Null Kuzushi*
3) *Contact Point Tsukuri*
4) *Rotatory Trajectory (easy to handle)*.
5) *Need high rotational strength*
6) *Easy Avoidance   but natural Renraku*
7) *Applied as psycological trick*